\documentclass[onecolumn,secnumarabic,amsmath,amssymb,balancelastpage,nofootinbib]{article}

\usepackage[e]{esvect}
\usepackage{xcolor}
\usepackage{graphics}      
\usepackage{graphicx}      
\usepackage{epsf}          
\usepackage{bm}            

\usepackage{natbib}
\usepackage{amssymb}
\usepackage{amsmath,esint}
\usepackage{mathrsfs}
\usepackage{framed}
\usepackage{bigints} 
\usepackage{enumitem}
\usepackage{soul} 
\usepackage{pifont}
\usepackage[capbesideposition={left,center},facing=yes,capbesidewidth=8cm,capbesidesep=quad]{floatrow} 
\usepackage{setspace} 

\usepackage{pdfcolfoot} 

\usepackage[none]{hyphenat} 

\usepackage[colorlinks=true]{hyperref}  

\usepackage[normalem]{ulem}

\definecolor{darkred}{rgb}{0.6,0,0}
\definecolor{darkgreen}{rgb}{0,0.5,0}
\definecolor{darkblue}{rgb}{0,0,0.6}
\definecolor{purple}{rgb}{0.5,0,0.6}
\hypersetup{ colorlinks,
linkcolor=darkblue,
filecolor=darkgreen,
urlcolor=darkgreen,
citecolor=darkred }

\begin{document}

\sloppy 

\bibliographystyle{authordate1}


\title{\Huge{Absorbing the Arrow of\\Electromagnetic Radiation}}
\author{Mario Hubert\\
Department of Philosophy\\The American University in Cairo
\\\\Charles T. Sebens\\Division of the Humanities and Social Sciences\\California Institute of Technology}
\date{April 2, 2023\\arXiv v.2\\\vspace{12 pt}Forthcoming in\\\emph{Studies in History and Philosophy of Science}}

\maketitle
\vspace*{0 pt}
\begin{abstract}
We argue that the asymmetry between diverging and converging electromagnetic waves is just one of many asymmetries in observed phenomena that can be explained by a past hypothesis and statistical postulate (together assigning probabilities to different states of matter and field in the early universe).  The arrow of electromagnetic radiation is thus absorbed into a broader account of temporal asymmetries in nature.  We give an accessible introduction to the problem of explaining the arrow of radiation and compare our preferred strategy for explaining the arrow to three alternatives: (i) modifying the laws of electromagnetism by adding a radiation condition requiring that electromagnetic fields always be attributable to past sources, (ii) removing electromagnetic fields and having particles interact directly with one another through retarded action-at-a-distance, (iii) adopting the Wheeler-Feynman approach and having particles interact directly through half-retarded half-advanced action-at-a-distance.  In addition to the asymmetry between diverging and converging waves, we also consider the related asymmetry of radiation reaction.
\end{abstract}

\newpage
\tableofcontents

\section{Introduction}\label{INTROsection}

\begin{figure}[htb]\centering
\includegraphics[width=9 cm]{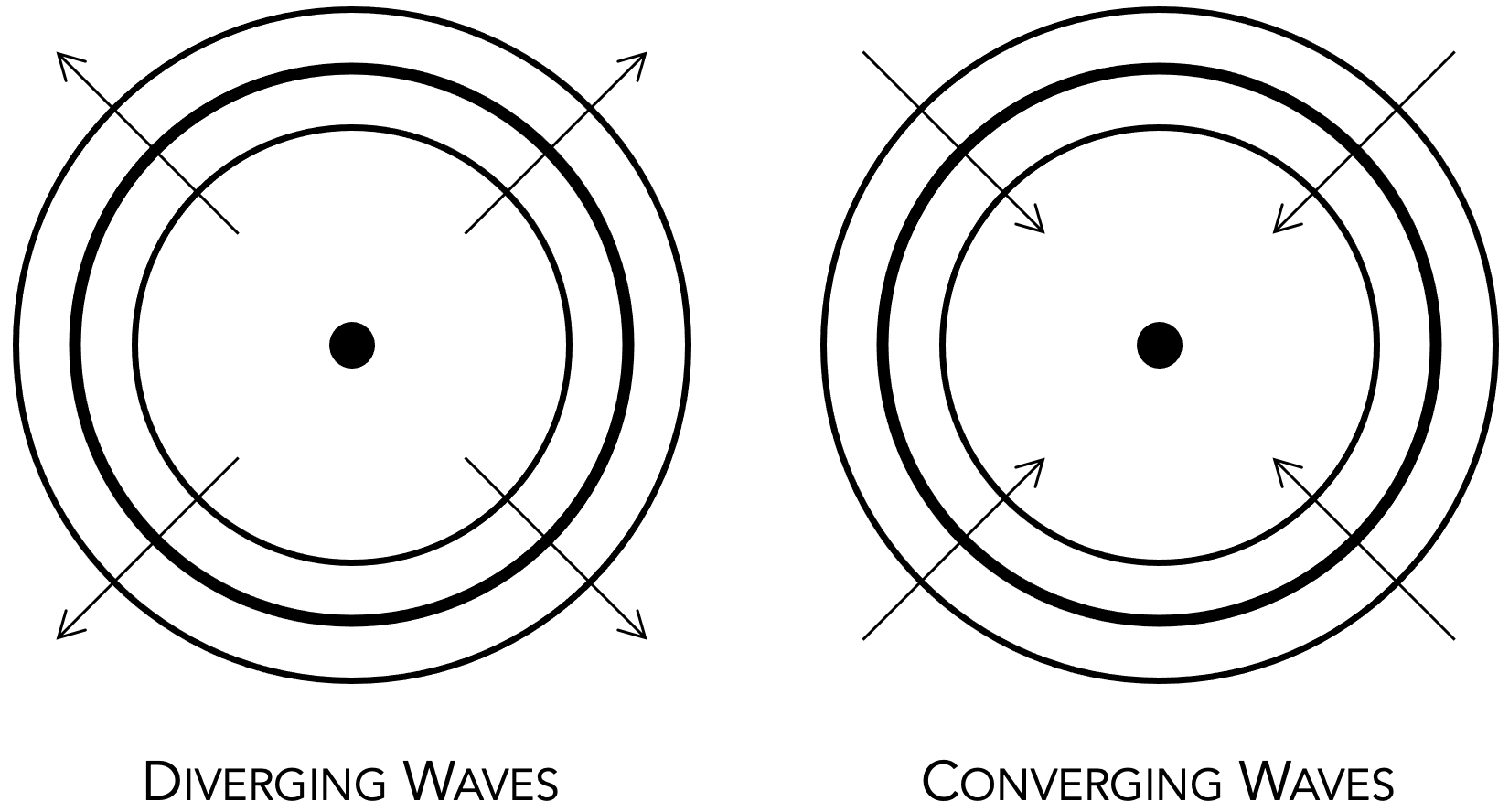}
\caption{The figure on the left shows diverging waves moving away from the center point and the figure on the right shows the time-reverse of this: converging waves approaching the center point.}
\label{wavefigure}
\end{figure}

The equations that describe water waves, sound waves, and electromagnetic waves are time-symmetric and allow for both diverging waves---that propagate outwards from a central point or region---and their time-reverse, converging waves (figure \ref{wavefigure}).  If you throw a stone in a pond, shout, or switch on a lightbulb, you will produce diverging waves (of water, sound, or light).  Diverging waves are commonplace.  Converging waves are allowed by the laws of the relevant physical theories, but in our world they are rare.  We never see circular waves spontaneously form from rustlings at the edge of a pond, increasing in amplitude as they converge towards the center.\footnote{See \citet{popper1956}; \citet[pg.\ 17]{zeh2007}.}  Still, converging waves can occur.  One way to make such waves would be to place a large floating ring on a calm body of water and carefully pulse it up and down (while keeping it level).\footnote{An example like this one appears in \citet[pg.\ 119]{davies}.}  This would result in converging waves within the ring and diverging waves outside of it.

There are a great many processes that, like diverging waves, can happen in reverse but hardly ever do.  Price has called the question as to why these processes occur in one temporal order but not the other ``the puzzle of temporal bias'':
\begin{quote}
``Late in the nineteenth century, physics noticed a puzzling conflict between the laws of physics and what actually happens.  The laws make no distinction between past and future---if they allow a process to happen one way, they allow it in reverse.  But, many familiar processes are in practice `irreversible,' common in one orientation but unknown `backwards.'  Air leaks out of a punctured tire, for example, but never leaks back in.  Hot drinks cool down to room temperature, but never spontaneously heat up.  Once we start looking, these examples are all around us---that's why films shown in reverse look so odd.  Hence the puzzle: What could be the source of this wide-spread temporal bias in the world, if the underlying laws are so even-handed?'' \citep[pg.\ 219]{price2004}
\end{quote}

Among philosophers of physics, there has emerged a fairly widespread consensus on how to solve the puzzle of temporal bias (though there is disagreement in the details), at least for standard thermodynamic processes like air leaking out of a tire or hot drinks cooling to room temperature.  We can explain why the reversed processes rarely occur by introducing a probability distribution over initial conditions that deems improbable the kind of fine-tuning that would be necessary for such reversed processes to be common.  \citet{albert2000} has given a clear and influential presentation of this kind of solution, calling the two posits that specify the probability distribution over initial conditions ``the past hypothesis'' and ``the statistical postulate.''  \citet{wallace2011} has written: ``There are no consensus positions in philosophy of statistical mechanics, but the position that David Albert eloquently defends in \emph{Time and Chance} \dots\ is about as close as we can get.''

In stark contrast (and despite considerable work on the subject), there has emerged no consensus as to how we ought to solve the puzzle of temporal bias for wave phenomena in general or for electromagnetic waves in particular.  There is no generally accepted explanation for the observed arrow of electromagnetic radiation.  One popular strategy\footnote{For a list of authors in addition to \citet{north2003} and \citet{atkinson2006} who take this option, see footnote \ref{longlist}.} is to give a statistical explanation of the arrow of radiation.  However, supporters of such an explanation vary considerably in the details and there are opponents defending quite different approaches.   In an effort to move the community towards consensus, here we are throwing our support behind \citeauthor{north2003}'s (\citeyear{north2003}) statistical strategy for explaining the arrow of radiation, identifying the work that remains to be done in developing the strategy, and comparing this strategy to its main rivals (focusing on the comparison to distant rivals over close ones).

\citet{north2003} builds on the broad agreement as to how we ought to explain thermodynamic asymmetries and argues that we can explain wave asymmetries using the same tools.  Converging waves (of sound, water, light, etc.) are rare because it would require extreme fine-tuning in the initial conditions for such waves to be common.  We can introduce a probability distribution over initial conditions for matter and field that makes such fine-tuning incredibly unlikely.  Because we believe that a single probability distribution over initial conditions will suffice to explain deflating tires, cooling beverages, diverging waves in general, and diverging electromagnetic waves in particular, we are seeking to ``absorb'' the arrow of electromagnetic radiation into a broader explanatory schema---unifying the arrow of radiation with other arrows of time.\footnote{The arrow of electromagnetic radiation is distinct from the arrow of time itself, if there is such a thing.  The arrow of radiation is about the time-directed nature of certain wave phenomena, similar to the arrow of entropy increase describing thermodynamic phenomena in our universe.
For \citet{Maudlin:2007ad}, time is intrinsically directed.  Nevertheless, one can still investigate the arrows of radiation and entropy increase with respect to this fundamental arrow of time.  Taking a different view, \citet{albert2000,Albert:2015aa} and \citet{Loewer:2012aa,Loewer:2012ab,Loewer:2020aa} deny the existence of a fundamental arrow of time and argue that the arrows of time we observe (like the arrow of entropy increase) can be explained without time itself being directed.  Of the four strategies for explaining the arrow of radiation explored here, two involve time-directed laws that seem to require a fundamental arrow of time (the Sommerfeld Radiation Condition approach and the retarded action-at-a-distance approach) and two do not require a fundamental arrow of time (the statistical approach that we favor and the Wheeler-Feynman approach).  In what follows, our focus will be on the arrow of radiation and not the arrow of time itself.}

This article begins with a section on technical background followed by a section presenting and defending the above-described statistical explanation of the arrow of radiation.  Then, we spend a section each on three competing strategies for explaining the arrow:  First, one can impose an additional time-asymmetric law (or postulate) that goes beyond Maxwell's equations in constraining the behavior of the electromagnetic field: the Sommerfeld Radiation Condition.  This condition requires that the electromagnetic field at any point in space and time be attributable to past sources.  Second, one can eliminate the electromagnetic field and have charges interact with one another directly over spatial and temporal gaps in a retarded action-at-a-distance theory, where the electromagnetic force on a given charge is determined by the past behavior of other charges (a move that was advocated by Walther Ritz in his 1909 debate with Albert Einstein).  Third, one can adopt the Wheeler-Feynman half-retarded half-advanced action-at-a-distance theory where the electromagnetic field is eliminated and the force on a charge is determined by both the past and the future behavior of other charges.

We include subsections evaluating, in detail, the strengths and weaknesses of each strategy for explaining the arrow of radiation.  To briefly summarize, here are a few advantages of the statistical strategy for explaining the arrow of radiation:  The statistical strategy does not require complicating the laws of electromagnetism through any addition or revision.  Unlike the three approaches just described, the statistical strategy gives a unified account of all wave and thermodynamic asymmetries.  The three alternative approaches draw a sharp distinction between matter and field that we believe to be unwarranted---treating the electromagnetic field as either unreal or as merely an emanation from charged matter.  (Debates over the right explanation of the arrow of radiation are thus tightly linked to debates over the ontological status of the electromagnetic field---debates that are important for the sake of better understanding both classical electromagnetism and quantum field theory.) The three alternative approaches do not entirely avoid statistical reasoning and, given that such reasoning will feature in any explanation of the arrow of radiation, we find a fully statistical explanation to be appealing.

In addition to the observed asymmetry between diverging and converging electromagnetic waves, there is a related asymmetry of radiation reaction.  When a charged body is accelerated from rest, it emits electromagnetic radiation and feels a radiation reaction force opposing the acceleration.  If we take charged matter to be composed of extended charge distributions, then the asymmetry of radiation reaction can be viewed as a consequence of the asymmetry of radiation.  As electromagnetic waves pass through an extended charged body on their way out, they exert a force on that body.  Radiation reaction is the result of self-interaction.  If we take charged matter to be composed of point charges, then in most versions of electromagnetism one will need to modify the Lorentz force law to account for radiation reaction (viewing the source of the radiation reaction asymmetry as distinct from the source of the radiation asymmetry).  The exception is Wheeler-Feynman electrodynamics, where for point charges the arrows of radiation reaction and radiation emission are both explained by the dynamical equations together with assumptions about an absorbing medium that surrounds all of the charges.

Throughout the article, we focus on classical electrodynamics.  This is standard practice in the literature on the arrow of radiation, though there are exceptions (\citealp{arntzenius1993, atkinson2006}).  One may wonder why we should try to explain the arrow of radiation within classical physics when we know that classical physics has been superseded by quantum physics and classical electromagnetism has been replaced by quantum electrodynamics.  This question is especially pressing given our discussions of the early universe and the ultimate composition of charged bodies.  We think it is important to see whether and how the arrows of radiation and radiation reaction can be explained within classical electromagnetism. Our methodology is to push the classical theory to its limits and see what it can do, recognizing that there will be places where further physics is needed.  Figuring out how the arrows of radiation and radiation reaction are best explained within classical electromagnetism provides insight into the laws and ontology of the theory.  Such work may also help us solve foundational problems in quantum electrodynamics.  In particular, studying self-interaction in classical electrodynamics can provide clues as to how self-interaction should be handled in quantum electrodynamics (an important, and notoriously difficult, subject).\footnote{See \citet{fundamentalityoffields}.}

This article is intended as an accessible entry point to debates about the arrow of radiation in classical electromagnetism and also as a comparative case for a particular explanation of this arrow.  Readers who are well-versed in the relevant literature may be particularly interested in the following highlights:  In section \ref{EMsection}, we differentiate our understanding of the arrow of radiation from characterizations of the arrow by Frisch and North.  That section closes with a discussion of the relation between the arrow of radiation and the arrow of radiation reaction for both extended charges and point charges, noting that the point charge Lorentz-Dirac equation breaks down if converging waves are present.  In section \ref{Ssection}, we discuss cosmic microwave background radiation and, unlike North, do not take it to be evidence for the existence of free (unsourced) electromagnetic fields.  We also depart from North by showing that backwards causation can be avoided in a statistical explanation of the arrow of radiation. In section \ref{SRCsection}, we present serious problems for formulating the Sommerfeld Radiation Condition (prohibiting unsourced electromagnetic fields) if there was a first moment and instead assume an infinite past.  In section \ref{WFsection} we separate out two absorber conditions in Wheeler-Feynman electromagnetism, noting that it is the second absorber condition that yields time-asymmetry.

\section{Waves in Classical Electromagnetism}\label{EMsection}

As background for the upcoming discussion of the arrow of electromagnetic radiation, let us briefly review some important features of classical electromagnetism.  At every moment, the magnetic field must be divergenceless,
\begin{equation}
\vec{\nabla}\cdot\vec{B}=0\ ,
\label{gaussB}
\end{equation}
and the divergence of the electric field must be proportional to the density of charged matter, $\rho$,
\begin{equation}
\vec{\nabla} \cdot \vec{E}=4\pi\rho\ .
\label{gaussE}
\end{equation}
These are Gauss's laws for electricity and magnetism, two of Maxwell's equations.  The time evolution of the electric and magnetic fields (the electromagnetic field) is determined by the remaining two of Maxwell's equations,
\begin{align}
\vec{\nabla} \times \vec{E}&=-\frac{1}{c}\frac{\partial \vec{B}}{\partial t}
\label{maxwellfaraday}
\\
\vec{\nabla} \times \vec{B}&=\frac{4\pi}{c}\vec{J}+\frac{1}{c}\frac{\partial \vec{E}}{\partial t}
\ ,
\label{ampere}
\end{align}
where $\vec{J}$ is the current density.  The time evolution of matter is given by a force law (such as the Lorentz force law or the Lorentz-Dirac force law) plus further equations governing other interactions (that lie outside of classical electromagnetism).  Solving all of these equations together is difficult.  In this section we will focus on the task of finding electric and magnetic fields that obey Maxwell's equations given a stipulated history for the charged matter.  We will model matter here as a continuous charge distribution, but one could derive equations for point charges as a special case.\footnote{The discussion in this section most closely follows that of \citet[ch.\ 10]{griffiths}, though the equations here are written in Gaussian cgs units.}

The electric and magnetic fields can be expressed in terms of the scalar potential $\phi$ and the vector potential $\vec{A}$ as
\begin{align}
\vec{E}&=-\vec{\nabla}\phi-\frac{1}{c}\frac{\partial \vec{A}}{\partial t}
\nonumber
\\
\vec{B}&=\vec{\nabla} \times \vec{A}
\ .
\label{fieldsfrompotentials}
\end{align}
Working with such potentials ensures that two of Maxwell's equations, \eqref{gaussB} and \eqref{maxwellfaraday}, will be automatically satisfied.  These potentials have a gauge freedom that can be partially fixed by adopting the Lorenz gauge condition,
\begin{equation}
\vec{\nabla}\cdot\vec{A}=-\frac{1}{c}\frac{\partial \phi}{\partial t}\ .
\end{equation}
With this condition in place, the remaining two Maxwell equations, \eqref{gaussE} and \eqref{ampere}, become
\begin{align}
\vec{\nabla} \cdot \vec{E}=4\pi\rho\quad&\Rightarrow\quad\left(\nabla^2-\frac{1}{c^2}\frac{\partial^2}{\partial t^2}\right)\phi=-4\pi\rho
\label{phiwave}
\\
\vec{\nabla} \times \vec{B}=\frac{4\pi}{c}\vec{J}+\frac{1}{c}\frac{\partial \vec{E}}{\partial t}\quad&\Rightarrow\quad\left(\nabla^2-\frac{1}{c^2}\frac{\partial^2}{\partial t^2}\right)\vec{A}=-\frac{4\pi}{c}\vec{J}\ ,
\label{awave}
\end{align}
These are wave equations for each potential.

\begin{figure}[!ht]\centering
\includegraphics[width=12 cm]{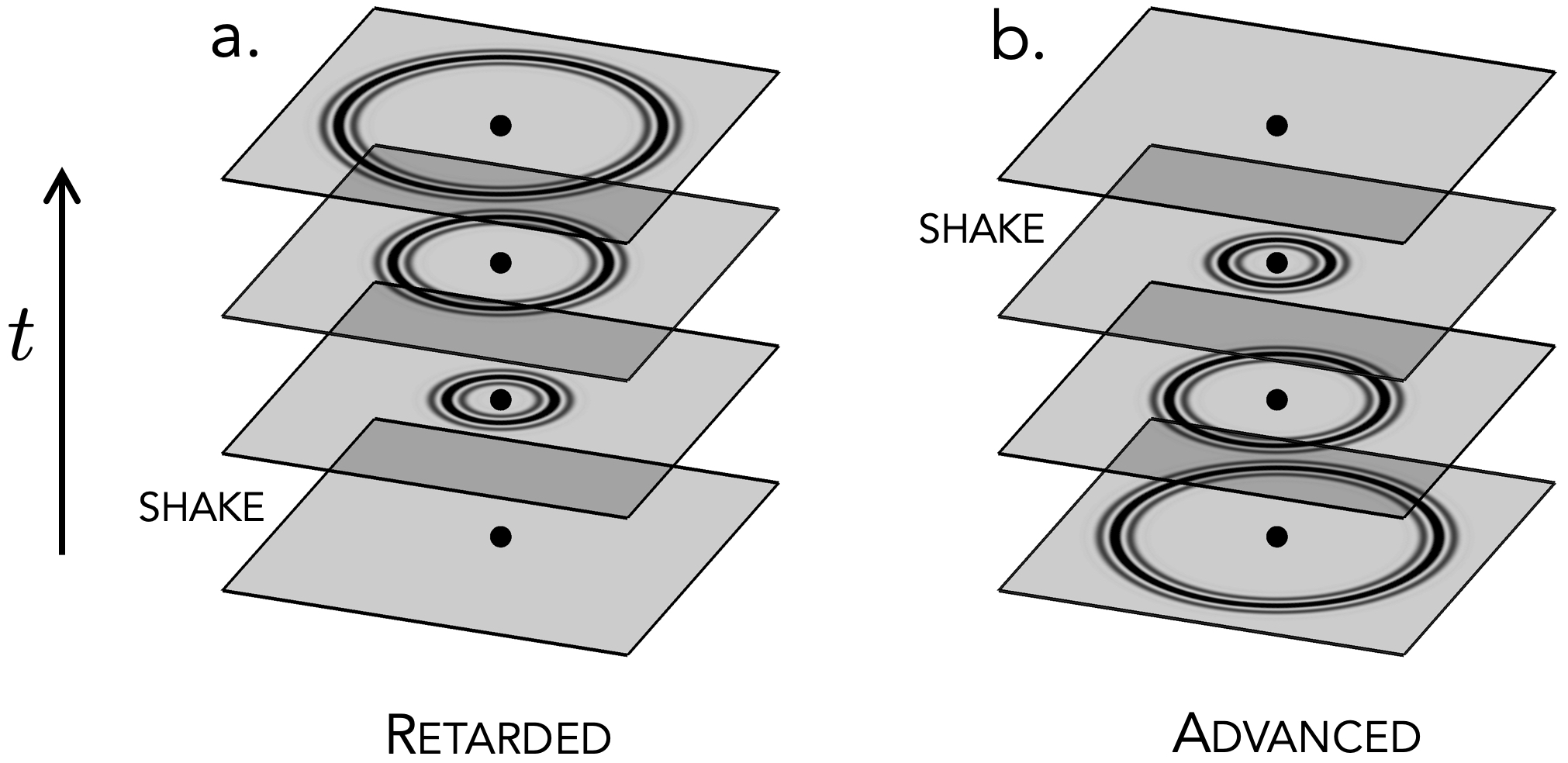}
\caption{On the left, a charge is shaken and sends out diverging waves (the retarded solution).  On the right, the shaken charge functions as a sink for converging waves (the advanced solution).  The motion of charged matter is the same in both figures, but the non-electromagnetic forces needed to account for that motion would be different (as energy is transferred from matter to field in figure a and from field to matter in figure b).  To simplify the depiction of electromagnetic waves, the dark rings only show the distribution of field energy.}
\label{retarded-advanced}
\end{figure}

The following expressions for $\phi$ and $\vec{A}$ satisfy both \eqref{phiwave} and \eqref{awave},
\begin{align}
\phi(\vec{x},t)&=\int d^3\vec{x}' \frac{\rho(\vec{x}',t_r)}{|\vec{u}|}
\nonumber
\\
\vec{A}(\vec{x},t)&=\frac{1}{c}\int d^3\vec{x}' \frac{\vec{J}(\vec{x}',t_r)}{|\vec{u}|}\ ,
\label{retardedsolutions}
\end{align}
where $\vec{u}$ is a vector that points from $\vec{x}'$ to $\vec{x}$, $\vec{u}=\vec{x}-\vec{x}'$, and $t_r$ is the retarded time, $t_r=t-\frac{|\vec{u}|}{c}$ (the time that a signal traveling at the speed of light from $\vec{x}'$ would have to have been emitted for it to arrive at $\vec{x}$ at $t$).  These are called the \emph{retarded} solutions of \eqref{phiwave} and \eqref{awave}.  The potentials at a point can be calculated by combining contributions to the field associated with bits of charged matter at distant points at past (retarded) times.  That is, one can find the values for $\phi$ and $\vec{A}$ at a given point by integrating contributions to these potentials from the charge and current densities at each point $\vec{x}'$ at the appropriate moment in the past, $t_r$.  You might interpret \eqref{retardedsolutions} as telling us how past charged matter acts as source for the current electromagnetic field.  For the simple case of a charge that is briefly shaken back and forth, the electromagnetic field calculated from the retarded potentials will describe diverging electromagnetic waves propagating outwards after the charge is shaken, carrying away energy (figure \ref{retarded-advanced}.a).

The solutions in \eqref{retardedsolutions} are not the only solutions to \eqref{phiwave} and \eqref{awave}.  There are also \emph{advanced} solutions,
\begin{align}
\phi(\vec{x},t)&=\int d^3\vec{x}' \frac{\rho(\vec{x}',t_a)}{|\vec{u}|}
\nonumber
\\
\vec{A}(\vec{x},t)&=\frac{1}{c}\int d^3\vec{x}' \frac{\vec{J}(\vec{x}',t_a)}{|\vec{u}|}\ ,
\label{advancedsolutions}
\end{align}
where the only difference from \eqref{retardedsolutions} is that the retarded time, $t_r$, is replaced by the advanced time, $t_a=t+\frac{|\vec{u}|}{c}$ (the time that a signal traveling at the speed of light from $\vec{x}$ at $t$ would arrive at $\vec{x}'$).  Using \eqref{advancedsolutions}, the potentials at a point can be calculated by combining contributions to the field associated with bits of charged matter at distant points at future (advanced) times.  You might interpret \eqref{advancedsolutions} as telling us how future charged matter acts as sink for the current electromagnetic field.  For a charge that is briefly shaken back and forth, the electromagnetic field calculated from the advanced potentials will describe converging electromagnetic waves propagating inwards, arriving at the charge as it is being shaken and depositing energy in the charge (figure \ref{retarded-advanced}.b).  For such an advanced solution, one might be tempted to say that the presence of converging electromagnetic waves at some time before the shaking is caused by the future shaking of the charge (that there is retrocausation).  In this article, we will avoid such language and generally work under the assumption that causes come before their effects.  One can retain the ordinary picture of causes preceding their effects in this case if one views the converging waves at a particular moment as caused by the earlier presence of more widely spread out converging waves and as causing the future motion of the charge (alongside other forces).    To avoid such waves that come in from the infinite past and are not produced by earlier motions of charged bodies,  one might stipulate that it is the retarded and not the advanced solutions that are to be used.  We will consider the merits of such a proposal in section \ref{SRCsection}.

In addition to the retarded and advanced solutions in \eqref{retardedsolutions} and \eqref{advancedsolutions}, there are also \emph{free} solutions which describe the propagation of electromagnetic waves in the absence of charges (when the right-hand sides of \eqref{phiwave} and \eqref{awave} are zero).  By the Kirchhoff representation theorem,\footnote{See \citet[sec.\ 2.3]{earman2011}.} an arbitrary solution to \eqref{phiwave} and \eqref{awave} can be written as the sum of the retarded solution $(\phi_{ret},\vec{A}_{ret})$ and a free solution $(\phi_{in},\vec{A}_{in})$ or, alternatively, as the sum of the advanced solution $(\phi_{adv},\vec{A}_{adv})$ and a different free solution $(\phi_{out},\vec{A}_{out})$.  The solution can be written either as
\begin{align}
\phi_{tot}(\vec{x},t)&=\phi_{ret}(\vec{x},t)+\phi_{in}(\vec{x},t)
\nonumber
\\
\vec{A}_{tot}(\vec{x},t)&=\vec{A}_{ret}(\vec{x},t)+\vec{A}_{in}(\vec{x},t)
\label{Rrepresentation}
\end{align}
or
\begin{align}
\phi_{tot}(\vec{x},t)&=\phi_{adv}(\vec{x},t)+\phi_{out}(\vec{x},t)
\nonumber
\\
\vec{A}_{tot}(\vec{x},t)&=\vec{A}_{adv}(\vec{x},t)+\vec{A}_{out}(\vec{x},t)
\label{Arepresentation}
\ .
\end{align}
The ``in'' in $(\phi_{in},\vec{A}_{in})$ stands for ``incoming,'' as this contribution to the potentials cannot be traced back to past charged matter sources and is thus thought of as coming in from the infinite past.  The ``out'' in $(\phi_{out},\vec{A}_{out})$ stands for ``outgoing,'' as this contribution to the potentials cannot be traced forward to future charged matter sinks and is thus thought of as going out to the infinite future.

Generalizing from these two specific ways of decomposing the field, one can write arbitrary potentials satisfying \eqref{phiwave} and \eqref{awave} as the sum of some constant $\alpha$ times the retarded solution plus $1-\alpha$ times the advanced solution plus a free solution (that is $\alpha$ times the incoming solution plus $1-\alpha$ times the outgoing solution):
\begin{align}
\phi_{tot}(\vec{x},t)&=\alpha\phi_{ret}(\vec{x},t)+(1-\alpha)\phi_{adv}(\vec{x},t)+\alpha\phi_{in}(\vec{x},t)+(1-\alpha)\phi_{out}(\vec{x},t)
\nonumber
\\
\vec{A}_{tot}(\vec{x},t)&=\alpha\vec{A}_{ret}(\vec{x},t)+(1-\alpha)\vec{A}_{adv}(\vec{x},t)+\alpha\vec{A}_{in}(\vec{x},t)+(1-\alpha)\vec{A}_{out}(\vec{x},t)
\ .
\label{Grepresentation}
\end{align}
In section \ref{WFsection}, we will discuss taking $\alpha$ to be one-half and eliminating the free field solution (the Wheeler-Feynman approach).

Corresponding to the retarded, advanced, incoming, and outgoing potentials, we can speak of the retarded, advanced, incoming, and outgoing electric and magnetic fields, using \eqref{fieldsfrompotentials} to pass from potentials to fields.  Or, we can combine the electric and magnetic fields into the Faraday tensor $F^{\mu \nu}$ and speak of retarded, advanced, incoming, and outgoing electromagnetic fields.  In discussions of the arrow of radiation, the tensor indices are often dropped and these fields are written as $F_{ret}$, $F_{adv}$, $F_{in}$ and $F_{out}$.  We will adopt this terse notation in future sections.\footnote{Although we used the scalar and vector potentials in the Lorenz gauge to pick out the retarded, advanced, incoming, and outgoing electromagnetic fields, these separate fields can be written using the potentials in any gauge or using the gauge-independent electric and magnetic fields (or the gauge-independent Faraday tensor).  There is no need to ontologically privilege the Lorenz gauge, though one may choose to do so (\citealp{Maudlin:2018aa} discusses the pros and cons).}

To better understand the Kirchhoff representation theorem, let us again consider the earlier example of a charge that is shaken and emits electromagnetic waves (figure \ref{retarded-advanced}.a).  In the retarded representation \eqref{Rrepresentation}, we have a retarded electromagnetic field describing diverging electromagnetic waves leaving the charge after it is shaken (and also the Coulomb field around the charge).  There is no incoming (free) field.  In the advanced representation \eqref{Arepresentation} of the very same history, we have an advanced electromagnetic field describing electromagnetic waves converging on the charge---reaching it when it shakes (and also the Coulomb field around the charge).  In addition to this advanced field, there is an outgoing (free) field.  Before the shaking, the outgoing field cancels the waves in the advanced field (destructive interference).  After the shaking, the outgoing field describes the very same diverging electromagnetic waves that the retarded field described in the retarded representation.\footnote{\citet[pg.\ 1089]{north2003} describes a similar case, considering the turning on of a light bulb in the advanced representation.}  Thus, diverging waves can be expressed using either retarded or advanced fields (with the appropriate free fields).  In this case, energy is emitted from charged matter into the electromagnetic field and we see that this energy emission can be described using either the retarded or advanced representation.  Similarly, cases of energy absorption can be described using either the retarded or advanced representation.\footnote{On the point that both representations are fully capable of describing energy emission and absorption, see \citet[pg.\ 1088]{north2003}; \citet[][pg.\ 141]{frisch2005}; \citet[][pg.\ 18]{zeh2007}.}

The asymmetry we seek to explain is the observed asymmetry between converging and diverging electromagnetic waves in the total electromagnetic field.  Why is it that symmetric converging waves are rare?  Of course, waves that increase in strength over time are not so rare.  Consider, for example, the way in which ocean waves converge at an uneven shoreline or the way in which ear trumpets concentrate sound waves to aid hearing.  In this article, when we speak of ``converging'' waves we mean waves that approach a central point or region in a symmetric, coordinated manner.  If you oscillate a charge up and down in the z direction, it will produce a diverging electromagnetic wave that is symmetric about the z axis.  The time-reverse of this process is a converging wave that is symmetric about the z axis.  We seek to explain why this kind of converging wave is so rare.  Choosing a particular representation does not give an explanation as to why converging waves are rare.  Choosing the retarded representation and stipulating that there are no incoming free fields does yield such an asymmetry (section \ref{SRCsection}).  However, we think there is a better way to explain the absence of converging waves: they are improbable.

There is disagreement in the literature as to the arrow of electromagnetic radiation that needs to be explained.\footnote{\citet{price1996, price2006} takes the arrow of radiation to be a macroscopic effect:
\begin{quote}
``According to this view the radiative asymmetry in the real world simply involves an imbalance between transmitters and receivers: large-scale sources of coherent radiation are common, but large receivers, or `sinks,' of coherent radiation are unknown. ... At the microscopic level things are symmetric, and we have both coherent sources and coherent sinks.  At the macroscopic level we only notice sources, however, because only they combine in an organized way in sufficiently large numbers.'' \citep[pg.\ 71]{price1996}
\end{quote}
We find it odd to draw such a sharp distinction between transmitters and receivers.  Really, there are just charges and charges sometimes emit and sometimes absorb energy.  Also, radiation reaction illustrates that there is time asymmetry at both the macro and micro level.  We thus think it is wrong to say that ``at the microscopic level things are symmetric'' (see \citealp[pg.\ 139--142]{frisch2005}, especially the point about synchrotron radiation).}  \citet[pg.\ 384]{frisch2000} initially sought to explain why the incoming free field is zero in the retarded representation, a formulation of the problem that we would resist because some proposed explanations of the arrow of radiation involve non-zero incoming free fields.  In his later book, \citet[pg.\ 108]{frisch2005} describes the asymmetry-to-be-explained as follows:
\begin{quote}
``There are many situations in which the total field can be represented as being approximately equal to the sum of the retarded fields associated with a small number of charges (but not as the sum of the advanced fields associated with these charges), and there are almost no situations in which the total field can be represented as being approximately equal to the sum of the advanced fields associated with a small number of charges.''
\end{quote}
This is a better formulation of the arrow, but it is a step removed from what we actually observe: diverging waves in the total electromagnetic field \citep[sec.\ 2.4]{price2006}.  \citet[sec.\ 2]{north2003} seeks to explain why ``accelerated charges produce retarded and not advanced radiation'' or, put more precisely, why ``the retarded solution requires a much more natural free-field component than the advanced solution.''  In particular, North takes the weak and relatively uniform cosmic microwave background (CMB) to be the incoming free field in the retarded representation.  As will be discussed in section \ref{FIELDsection}, we do not want to assume that the CMB is a truly free field or that the true incoming free field is simple in the retarded representation.  Such assumptions could be part of an explanation of the arrow of radiation, but they are not part of the arrow itself.

Before embarking on the project of explaining the arrow of radiation, we should pause to discuss both the time-reversal invariance of electromagnetism and the asymmetry of radiation reaction.  First, let us briefly consider the question as to whether the laws of electromagnetism are time-symmetric (or, put another way, whether they are time-reversal invariant).  Recall that the puzzle of temporal bias for electromagnetic waves (posed in section \ref{INTROsection}) asks why we observe waves behaving in a time-directed way (diverging but not converging) when the underlying laws are time-symmetric.  In fact, there is debate as to whether the laws of electromagnetism are truly time-symmetric.\footnote{See \citet[ch.\ 1]{albert2000}; \citet{malament2004, arntzenius2009, allori2015, struyve2020, roberts2021}.}  If you simply reverse the order of instantaneous states for matter and field without altering the electric or magnetic fields at each moment, then the time-reverse of a history obeying Maxwell's equations will, in general, not obey Maxwell's equations.  If, instead, you reverse the order of states and flip the orientation of the magnetic field, then the time-reverse of a history obeying Maxwell's equations will obey Maxwell's equations.  Applying this second form of time reversal, the time-reverse of a history where the electromagnetic field is purely retarded (with no incoming field) is a history where the electromagnetic field is purely advanced (with no outgoing field).  The time-reverse of a series of diverging waves is a series of converging waves.  Whether or not we count this form of time-reversal as true time-reversal, it is sufficient to get the puzzle off the ground: If for every history with diverging waves there is a corresponding history with converging waves, why do we never see converging electromagnetic waves?

In addition to the asymmetry between diverging and converging electromagnetic waves, there is a related asymmetry in electromagnetic phenomena that needs to be explained: radiation reaction.  This asymmetry will not be our main focus, but we will discuss whether different proposals for explaining the wave asymmetry can also explain the asymmetry of radiation reaction.  Here is the asymmetry:  When a charged body accelerates, it emits radiation.  That radiation carries energy and momentum.  Because momentum is conserved, the change in field momentum is balanced by a change in momentum of the charged body (or whatever is accelerating it).  This radiation reaction is a time-asymmetric phenomenon somewhat similar to friction.  An accelerating charge will feel a reactive damping force that it would not feel if it were uncharged.  The time reverse of this effect would be an anti-damping force where the field gives energy to the charge and helps it accelerate.  That is not observed in nature.

For extended charged bodies, we can explain radiation reaction by analyzing the way that electromagnetic waves propagate through bodies on their way out.\footnote{See \citet{rohrlich1999, rohrlich2000}; \citet[sec.\ 2.2]{gravitationalfield}.  Note that the forces exerted by the electromagnetic field within an accelerating extended charged body balance both the energy lost to radiation and the energy transferred from the body to the electromagnetic field that surrounds it and travels along with it.}  If we can explain why electromagnetic waves diverge, we can explain radiation reaction.  For point charges, the situation is more complicated.  Calculating the force on a point charge in classical electromagnetism is problematic because the electromagnetic field becomes infinitely strong as one approaches the charge (and the value of the field is undefined at the location of the charge).  One could try stipulating that point charges do not notice their own fields and only experience the standard forces from the fields of other particles (via the Lorentz force law, $\vec{F}=q\vec{E}+\frac{q}{c}\vec{v}\times\vec{B}$), but then one would miss radiation reaction and have violations of both energy and momentum conservation.

One way out of these troubles is to argue that there are no point charges in nature.\footnote{For discussion of this idea, see \citet[sec.\ 3]{arntzenius1993}; \citet[ch.\ 16]{jackson}; \citet[pg.\ 55--58, 117--118]{frisch2005}; \citet[pg.\ 145]{pietsch2012}; \citet[sec.\ 8]{positrons}; \citet[sec.\ 1.4]{wald2022}; \citet{fundamentalityoffields}.}  Should one desire to work with point charges, there are a variety of strategies available for patching up classical electromagnetism.  For example, one might replace the Lorentz force law with the Lorentz-Dirac force law,\footnote{See \citet{Dirac:1938aa}; \citet{Wheeler:1945aa}; \citet[sec.\ 3.3]{frisch2005}; \citet[sec.\ 3]{earman2011}; \citet{Kiessling:2011ab}; \citet[sec.\ 3.1]{lazarovici2018}.} which can be written in outline as
\begin{equation}
\vec{F}=\vec{F}_{ext}+\vec{F}_{rad}+\vec{F}_{inc}
\ .
\label{lorentzdirac}
\end{equation}
The first term, $\vec{F}_{ext}$, is the Lorentz force from the retarded electromagnetic fields associated with each of the other charges.  The second term, $\vec{F}_{rad}$, gives the (time-asymmetric) radiation reaction force on the charge.  This radiation reaction force can be expressed in terms of time derivatives of the charge's position without referencing any electromagnetic fields.  The final term, $\vec{F}_{inc}$, is the Lorentz force from the free incoming electromagnetic field: $\vec{F}_{inc}=q\vec{E}_{inc}+\frac{q}{c}\vec{v}\times\vec{B}_{inc}$.  For some free incoming fields, this force will be well-defined.  However, if the incoming field contains electromagnetic waves that converge on the charge, then this force will not be well-defined.  For example, consider the purely advanced solution in figure \ref{retarded-advanced}.b, where the total electromagnetic field is a wave that converges on a charge (which we will assume here to be a point charge).  In the retarded representation \eqref{Rrepresentation}, the incoming free field would be ill-defined at the location of the point charge when the wave converges.  Thus, the Lorentz-Dirac equation can break down.  In general, the time-reverse of a history of charged particles and fields that obeys the Lorentz-Dirac equation will be a history where the Lorentz-Dirac equation breaks down because $\vec{F}_{inc}$ is not always well-defined at the particle locations.  For the law to function properly, we must explain why the problematic incoming fields do not occur in nature.  An explanation as to why electromagnetic waves diverge might also explain this.  We will come back to this point in future sections.  To review, the Lorentz-Dirac equation for point charges is time-asymmetric but should not be viewed as the sole source of time-asymmetry in electromagnetism because its applicability presupposes time-asymmetric radiation.  There is more to be said about the strengths and weaknesses of the Lorentz-Dirac equation, but let us not go too deep into the problems of point charges.

\section{Strategy 1: A Statistical Explanation}\label{Ssection}

The motivating idea behind the statistical strategy for explaining the arrow of electromagnetic radiation is stated clearly by \citet{frisch2015}: converging waves are rare because ``a converging wave would require the co-ordinated behaviour of `wavelets' coming in from multiple different directions of space---delicately co-ordinated behaviour so improbable that it would strike us as nearly miraculous.''\footnote{In a similar complaint about the miraculous nature of converging waves, \citet{popper1958} writes: ``If not steered by an expanding wave, the contracting wave, though not in itself physically impossible, would nevertheless have the character of a physical miracle: it would be like a conspiracy, undertaken by many people, each carefully acting so as to support all the others, but without any previous arrangement, or anything like a prepared plan.''  Of course, converging waves are possible when you have a prepared plan.  With the right electromagnetic wave emitters arranged in a ring and set on a timer, you could get electromagnetic waves propagating inwards toward a central point.  (A similar example with water waves was given in the introduction.)}  As an example, consider the advanced solution in the case of a single charge briefly shaken (figure \ref{retarded-advanced}.b).  There we have electromagnetic waves that approach the charge from all sides in a coordinated manner that seems improbable.  To justify the claim that such histories are improbable, we need to say more about the probabilities for different histories.  \citet[pg.\ 18]{zeh2007} criticizes this kind of statistical explanation, writing:
\begin{quote}
``The popular argument that advanced fields are not found in Nature because they would require improbable initial correlations is known from statistical mechanics, but totally insufficient \dots\ The observed retarded phenomena are precisely as improbable among \emph{all possible} ones, since they describe equally improbable \emph{final} correlations.''
\end{quote}
Note that (by the Kirchhoff representation theorem) we cannot actually say that advanced fields are absent in nature because the electromagnetic field can always be decomposed into an advanced field and an outgoing (free) field \eqref{Arepresentation}.  What must be explained is the fact that converging electromagnetic waves are rarely found in nature.  We can break the symmetry that Zeh identifies by adding a statistical postulate assigning probabilities over initial conditions, not final conditions.  This is exactly the same move that is standardly made when one uses statistical mechanics to explain the observed asymmetries of thermodynamics.  Let us take a moment to review the philosophical foundations of statistical mechanics before returning to the arrow of electromagnetic radiation.

\subsection{The Past Hypothesis and the Statistical Postulate}\label{PHsection}

\begin{figure}[!ht]\centering
\includegraphics[width=5 cm]{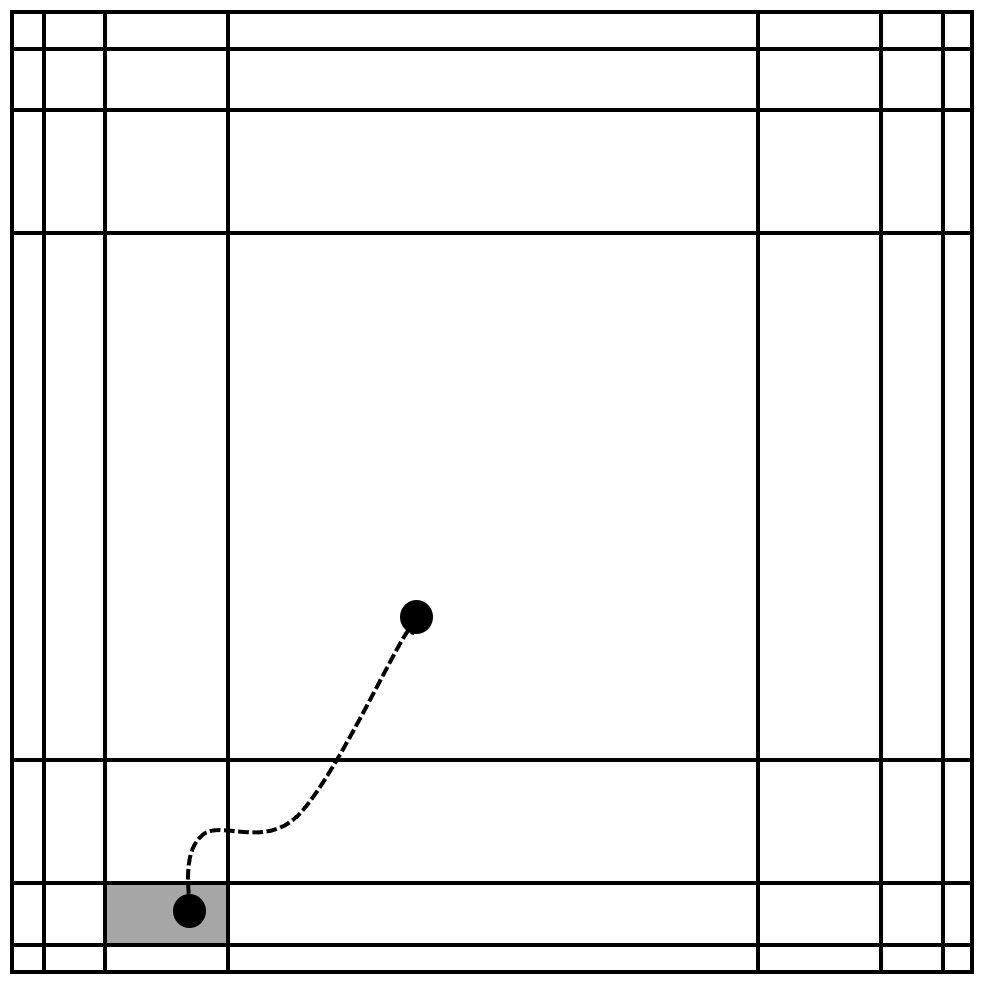}
\caption{This figure gives a depiction of the space of all accessible microstates, with small low-entropy microstates in the corners and a large maximum-entropy equilibrium macrostate in the center.  The point representing the microstate begins in the medium-sized gray macrostate and its evolution illustrates the second law of thermodynamics, moving through regions of increasing entropy until the system reaches thermal equilibrium.  This is exactly the evolution one would expect from the structure of this space.  Although it is not apparent in this simplified image, in reality the space is high-dimensional, and the vast majority of the many ways out of a small macrostate take you to a larger macrostate.\protect\setcounter{footnote}{17}\protect\footnotemark}
\label{fig:phase-space}
\end{figure}

 We will follow the current trend among philosophers of adopting a ``Boltzmannian'' or ``neo-Boltzmannian'' (as opposed to a ``Gibbsian'') approach to statistical mechanics.\footnote{For an introduction to this Boltzmannian approach and a comparison to the Gibbsian alternative, see \citet{callender1999}; \citet{albert2000}; \citet{goldstein2001}; \citet[sec.\ 4 and 5]{uffink2007}; \citet{frigg2008}; \citet[ch.\ 8]{carroll2010}; \citet{north2011}; \citet{wallace2015}; \citet{friggwerndl2019}; \citet{goldstein2020}; \citet[ch.\ 7]{Myrvold:2021tx}.}  On this approach, we can take the entropy $S$ of a system in a particular microstate to be proportional to the natural logarithm of the volume $W$ of the system's macrostate in the space of all accessible microstates,
\begin{equation}
\label{eq:entropy}
S=k\, \ln  W
\ ,
\end{equation}
where $k$ is Boltzmann's constant.  The previous sentence requires some unpacking.  For a monatomic gas in a box, the microstate is specified precisely by microscopic variables: the positions and velocities of all the atoms in the gas.  The macrostate is specified imprecisely by macroscopic variables (macrovariables) like the pressure, temperature, and volume of the gas.  The space of all accessible microstates for a closed system is an energy hypersurface within phase space.  Phase space is a $6N$-dimensional space with dimensions for the $x$, $y$, and $z$ components of the position and velocity of each of $N$ atoms.  By imposing constraints like the boundaries of the box, the total energy of the gas, and that the walls of the box do not transfer energy to the environment, we arrive at a $6N\!-\!1$-dimensional constant-energy hypersurface that is the accessible subspace within phase space. The macrostate picks out a region of this energy hypersurface and the microstate singles out a particular point in that region (see figure \ref{fig:phase-space}).

\footnotetext{Tim Maudlin has emphasized this point in his talk, ``Boltzmann Entropy, the Second Law, and the Architecture of Hell.''}

The motion of the particular point representing the gas is determined by the laws governing the collisions of atoms, which we might model as repulsive Newtonian forces that depend on the distances between the atoms.  The laws for collisions are ordinarily taken to be time-reversal-invariant, such that any sequence of events allowed by the laws is also allowed to occur in the opposite order.  However, the behaviors we observe are time-directed.  (This was the puzzle of temporal bias mentioned in the introduction.)  For example, a gas confined to the left half of a box will expand to fill the entire box if the barrier is removed.  The reverse process of a gas contracting to fill only half of a box never occurs.  This is just one instance of the Second Law of Thermodynamics. In one of its formulations, it says the following:
\begin{quote}
\textbf{The Second Law of Thermodynamics:}  The entropy of a closed system will almost always either increase or remain the same.
\end{quote}
Applying the definition of entropy in \eqref{eq:entropy}, this means that the point in phase space representing a gas will move into macrostates of equal or greater volume until it reaches the equilibrium macrostate (figure \ref{fig:phase-space}).

To explain the gas's time-directed behavior, we can apply a statistical postulate over the initial conditions at the moment the barrier is removed---assigning a uniform probability distribution over the region of the energy hypersurface compatible with the known values for the (appropriate) macrovariables.  According to this probability distribution, it is overwhelmingly likely that the gas's microstate will move through ever larger macrostates until it reaches equilibrium (and the gas is spread evenly throughout the entire box).  Motion into a smaller macrostate is physically possible but very unlikely.\footnote{The Boltzmann equation mathematically describes how the density of the gas changes in the box. Boltzmann derived this equation by making a statistical assumption about the collisions of the gas molecules, dubbed the \emph{Stoßzahlansatz}, which postulates that the gas molecules are typically uncorrelated when the barrier is removed. According to the Boltzmann equation, it is overwhelmingly likely that the gas will fill the box once the barrier is removed \citep{Brown:2009aa}.}

Imposing this kind of statistical postulate at the beginning explains why we get time-directed behavior (obeying the Second Law of Thermodynamics) afterwards.  But, such a postulate makes poor predictions about the past.  To see why, consider applying this kind of postulate at a time after the barrier is removed but before equilibration.  At this moment, one might include macrovariables that describe the unequal pressures (or densities) in the left and right halves of the box.  When we consider the positions and velocities for atoms that are consistent with this macrostate and assign a uniform probability distribution over the microstates compatible with the macrostate, the forward evolution will be exceedingly likely to fill the box.  So will the backwards evolution.  Knowing only the macrostate, one would not predict the gas to have been further concentrated in the left half of the box at earlier times (though in reality it was).

In general, if you apply the above kind of statistical postulate to a particular system at a particular time, you will predict that entropy will increase (or stay the same) both forwards and backwards in time.  To generate correct predictions for some time period of interest, you should apply a statistical postulate at the beginning of the time period.  If we are interested in everything that has happened in the history of our universe, we can apply such a statistical postulate sometime soon after the big bang:
\begin{quote}
\textbf{The Statistical Postulate:}  For the purposes of making predictions in the future of some time $t_0$ soon after the big bang, we should apply a uniform probability distribution to microstates compatible with the macrostate at $t_0$---where that macrostate is a region of the relevant energy hypersurface in the space of all possible microstates (phase space, or some successor to it) specified using appropriate macrovariables.
\end{quote}
This statement resembles the formulation in \citep[pg.\ 96]{albert2000}, though we are focusing here only on a single early moment (as in \citealp[][]{Loewer:2020aa}) and not directly specifying the probability distributions to be used for subsystems at later times.  The parenthetical about phase space allows for a revision of the degrees of freedom available to a system when we move to a physical description that includes more than just the positions and velocities of bodies---such as quantum theories\footnote{In quantum physics one can separately impose a Past Hypothesis and Statistical Postulate for the initial wave function or one can instead posit a particular density matrix at an early time \citep{chen2020}.} and, as we will see shortly, theories with fields.  The choice of ``appropriate'' macrovariables is left unspecified.

The Statistical Postulate will generate different predictions depending on the macrostate that is posited at $t_0$.  If the universe were in equilibrium at $t_0$, we would expect it to stay in (or near) equilibrium.  To generate accurate predictions, we can posit a low-entropy state:
\begin{quote}
\textbf{The Past Hypothesis:}  At some time $t_0$ soon after the big bang, the universe was in a particular low-entropy macrostate ``that the normal inferential procedures of cosmology will eventually present to us'' \citep[pg.\ 96]{albert2000}.
\end{quote}
Putting the Past Hypothesis and Statistical Postulate together, we have ``a probability map of the universe'' \citep{Loewer:2020aa} assigning probabilities to all possible initial states and thus to all possible histories of the universe.  Using this probability distribution, we can predict that systems will behave in a time-directed manner (obeying the Second Law of Thermodynamics) even if the underlying dynamical laws are time-symmetric.  We thus have a resolution of the puzzle of temporal bias, at least for certain phenomena.  Soon, we will see that this explanatory schema works for waves as well.

At this point, one might naturally wonder about the nature of the probabilities specified by the Past Hypothesis and the Statistical Postulate.  What exactly are we saying when we specify certain probabilities over initial conditions for the universe?  This is a reasonable point of concern, but not one that we will address here (see \citealp{allori2020}).  The interpretation of the probabilities will eventually need to be settled to complete the Boltzmannian approach and our soon-to-come application of this approach to explaining the arrow of radiation.

For the goal of making accurate predictions about thermodynamic processes, many different probability distributions would work just as well as the uniform one employed by the Statistical Postulate \citep{wallace2011}.  Thus, although we would like to defend a statistical explanation as to why electromagnetic waves diverge, we are not committed to the specific probability distribution given by the combination of the Past Hypothesis and the Statistical Postulate.\footnote{One idea for avoiding unjustified precision is to collect all of the admissible probability distributions, form an equivalence class, and reformulate the Statistical Postulate with this equivalence class of probability distributions. (It is debated whether equivalence classes of probability distributions capture the right degree of precision. \citealp{Rinard:2021vb} argues in a different context on imprecise probabilities that these equivalence classes are still too precise, \citealp[see also][p.\ 5]{Chen:2021uw} for this argument.)  Going this route, we would no longer be interested in the exact probability distribution over initial conditions but rather in something more coarse-grained: what kinds of initial conditions are overwhelmingly likely (or ``typical'') and what kinds are overwhelmingly unlikely (or ``atypical'').  One could show that typical initial conditions yield the usual thermodynamic asymmetries.  This approach to the statistical postulate has been dubbed the \emph{typicality account} \citep[see, for instance,][who defend this account]{goldstein2001, goldstein2012, Maudlin:2019ab,Hubert:2020aa}. For our purposes, we will stick to the ordinary statistical postulate given above.  One could easily adapt our soon-to-be-given explanation of the arrow of electromagnetic radiation by using a modified statistical postulate if one wished to fold our explanation into a typicality account.}

\subsection{Incorporating the Electromagnetic Field}\label{FIELDsection}

As formulated above, the Past Hypothesis and the Statistical Postulate do not explicitly mention the state of the electromagnetic field.  But, a full specification of the microstate of the universe at $t_0$ would require specifying the state of the electromagnetic field.  An accurate description of the state of the electromagnetic field at this time in the history of the universe would have to be quantum field theoretic.\footnote{See \citet[pg.\ 1095--1096]{north2003}; \citet[pg.\ 539]{atkinson2006}.}  For our purposes here, we will stick to classical physics (following the plan set out in the introduction).  We expect the lesson that converging electromagnetic waves are deemed improbable by appropriate versions of the Past Hypothesis and Statistical Postulate to be retained in a quantum treatment of the early universe.

We can use the simplified context of classical electromagnetism to give a fictional account of the early universe that illustrates the way in which the arrow of electromagnetic radiation can be explained statistically, noting that the details of that explanation will change as one moves to more advanced physics:  Long ago (at $t_0$),\footnote{You might think of this fictional time as around 400,000 years after the big bang \citep[appendix A]{hartle2005}, when charged matter formed a plasma in thermal equilibrium with the electromagnetic field (though the entire universe was not in equilibrium, as can be inferred from the potential for further expansion).} charged matter lived in a bath of electromagnetic radiation so intense and chaotic that, by inspection, one would not be able to discern clear converging or diverging waves.  At this time, there was no arrow of radiation in the phenomena to be explained.  It is here that we can apply the Statistical Postulate, adopting a uniform probability distribution over microstates compatible with the macrostate for matter and field (following \citealp{north2003} and \citealp{atkinson2006}\footnote{Other authors have defended broadly similar explanations of the arrow of radiation.  O.\ \citet{penrosepercival}  introduce a ``law of conditional independence'' saying that you could not have distant parts of the universe coordinate to form a converging wave because those parts of the universe were never in causal contact (see also O.\ \citealp{Penrose:2001aa}).  R.\ \citet{penrose1979} gives a cosmological explanation of the arrow of electromagnetic radiation, viewing that arrow and the thermodynamic arrows as all explained by a low-entropy initial state (presenting a version of the Past Hypothesis that is explicit about the low gravitational entropy in the early universe).  \citet[appendix A]{hartle2005} similarly employs a version of the Past Hypothesis to explain the arrow of radiation and the thermodynamic arrows of time, describing the radiation that was present in the early universe as lacking the kind of correlations that would give rise to converging waves. \citet[pg.\ 524]{earman2011} ends his article with a conjecture that ``any [electromagnetic] asymmetry that is clean and pervasive enough to merit promotion to an arrow of time is enslaved to either the cosmological arrow or the same source that grounds [the] thermodynamic arrow (or a combination of both).''  \citet[pg.\ 30]{arntzenius1993} seeks a unified explanation of the arrow of radiation and other ``arrows of time'' within quantum field theory: ``For a simpleminded philosopher like me, it would seem most satisfactory if a unified account could be given of all arrows of time. ... I have the hope of a unified statistical account within quantum field theory of all arrows of time.''  \citet[sec.\ 2.2]{zeh2007} gives a cosmological explanation of the arrow of radiation (criticized in \citealp[sec.\ 4]{frisch2000}) that is not statistical, describing the early universe as an ideal absorber with properties that allow us to ignore any free (incoming) fields that might have preceded it.  Although we would like to be able to claim Einstein as an ally, he does not consistently defend a statistical explanation of the arrow of radiation.  \citet{ritzeinstein1909} write ``Einstein believes that irreversibility is exclusively due to reasons of probability'' \citep{ritzeinstein1990}, but elsewhere \citet{einstein1909} concludes that ``The elementary process of the emission of light is, thus, not reversible'' \citep[pg.\ 112]{frisch2005}. (For more on Einstein's views, see \citealp[pg.\ 109--114]{frisch2005}; \citealp{frischpietsch2016}.)  Our goal here is not to focus on the subtle differences between the accounts of the authors just listed, but instead to present a strong version of the statistical approach that they are all clustering around (so that we can compare it to the very different approaches in section \ref{SRCsection}--\ref{WFsection}).\label{longlist}}).  As the universe expanded, we were left with charged matter in a very weak bath of radiation (the cosmic microwave background, CMB).  Although a weak bath of radiation could conceivably contain converging waves that are destined to grow in strength as they approach charges, the Statistical Postulate applied to the earlier strong bath makes the presence of any such waves exceedingly unlikely.  Converging waves would require precise coordination of electromagnetic field values at distant locations.  Such fine-tuned initial conditions are implausible, and rightly ruled improbable by the Statistical Postulate.

As \citet[pg.\ 1088, 1091]{north2003} tells the story, the cosmic microwave background radiation is composed of free fields---incoming fields that would have to be added to the retarded fields to get the total electromagnetic field at some point now.  Although for practical purposes it is reasonable to treat the cosmic microwave background radiation as free when analyzing the behavior of subsystems long after the big bang, there is no clear way to settle whether any of that radiation is truly free.\footnote{On this point, \citet[pg.\ 159]{lazarovici2018} (who opposes free fields) writes ``We will never be able to determine that some observed radiation is truly source-free, coming in `from infinity'. In fact, good scientific practice is to assume that it is \emph{not} and look for---or simply infer---the existence of material sources.''  (See also \citealp[ch.\ 2]{zeh2007}; \citealp[sec.\ 7.1]{pietsch2012}; \citealp[pg.\ 9]{wald2022}.)}  Within classical electromagnetism, figuring out whether there are any truly free fields (in the retarded representation) would require determining whether the history of charged matter along the entire infinite past light-cone of each point in space fully specifies the value of the electromagnetic field at that point, via \eqref{retardedsolutions}, or whether a free incoming field would have to be added, as in \eqref{Rrepresentation}.  In our actual universe, tracing that light-cone into the distant past will eventually take us back to times in the early universe where classical electromagnetism does not accurately approximate what is happening.  One could attempt to sort the quantum description of the electromagnetic field into a contribution attributable to past sources and a source-free contribution, but our limited knowledge of the early universe makes it hard to see how we could gain knowledge as to whether there is a non-zero source-free contribution.  Given the difficulty of ascertaining such a thing, we will not argue that there is an empirical case to be made for the kind of statistical approach to explaining the arrow of electromagnetic radiation outlined here, as compared to the alternative approaches to be discussed in the following sections (that do not allow for the possibility of free fields).  There are multiple ways to explain the observed absence of converging electromagnetic waves.  As we cannot settle the matter with data, we must look to other considerations.

\subsection{Evaluation}\label{SEVALsection}

We find the above brief account as to the origin of the arrow of radiation to be attractive for four main reasons (all mentioned in \citealp{north2003}). First, there is no need to modify the standard laws of electromagnetism to explain the arrow of radiation.  Some view versions of the Past Hypothesis and the Statistical Postulate as together forming an additional law (or pair of laws) (\citealp{Chen:2020tv}, \citeyear{chenF}; \citealp{Loewer:2020aa}).  If the initial probability distribution specified by the Past Hypothesis and Statistical Postulate is a law, then it is a law we already need to account for other asymmetries and not an additional law peculiar to this strategy for explaining the arrow of radiation.  One might consider the Lorentz-Dirac equation \eqref{lorentzdirac} to be a modification to the standard laws of electromagnetism.  This equation is not needed to explain the prevalence of diverging waves (the arrow of radiation) or to explain radiation reaction for extended charges, but, as will be discussed below, the Lorentz-Dirac force law can be adopted to explain radiation reaction for point charges.  Some such modification would be needed to explain the arrow of radiation reaction for point charges within any of the approaches to explaining the arrow of radiation presented here except for the Wheeler-Feynman approach (see section \ref{WFsection}).

Second, the statistical explanation gives a unified account of all wave asymmetries.  In general, converging waves are improbable because the strange initial conditions needed for them to occur are improbable.  \citet[pg.\ 119]{davies} explains this well for the case of waves in a pond,
\begin{quote}
``\dots\ waves on real ponds are usually damped away at the edges by frictional effects.  The reverse process, in which the spontaneous motion of the particles at the edges combine favourably to bring about the generation of a disturbance is overwhelmingly improbable, though not impossible, on thermodynamic grounds.''
\end{quote}
The fact that such coordination is improbable follows from the Statistical Postulate.  This postulate also explains why converging electromagnetic waves are improbable.

Third, the statistical explanation unifies wave asymmetries with familiar thermodynamic asymmetries---gases expand, ice cubes melt, etc.  These are two kinds of asymmetries that one might have expected would receive different explanations.  In fact, all of these  asymmetries follow from the Past Hypothesis and the Statistical Postulate, provided we include the electromagnetic field in our descriptions of microstates and macrostates.  The same probability distribution explains both why waves diverge and why entropy increases.

Fourth, the symmetric treatment of charged matter and electromagnetic field fits well with quantum field theory where charged matter and the electromagnetic field are modeled by very similar equations (suggesting that they are the same kind of thing---see \citealp{fundamentalityoffields}).  In the statistical approach advocated here, the field is just as real as matter and it has independent degrees of freedom (its state is not fixed by the behavior of charged matter, as in section \ref{SRCsection}, though it is constrained by it).  The appeal of such a picture has been expressed in a memorable way by \citet[pg.\ 590]{penrose1979}\footnote{See also \citet[pg.\ 196]{rohrlich}; \citet[pg.\ 141--142]{pietsch2012}.} while criticizing the Wheeler-Feynman approach (which we will come to in section \ref{WFsection}),
\begin{quote}
``And I have to confess to being rather out of sympathy with the whole [Wheeler-Feynman] programme, which strikes me as being unfairly biased against the poor photon, not allowing it the degrees of freedom admitted to all massive particles!''
\end{quote}
\citet[pg.\ 2, 9--10]{wald2022} makes a similar remark when he addresses the “pernicious myth” that electromagnetic fields are produced by charged matter (as in section \ref{SRCsection}),
\begin{quote}
``\dots the view that electromagnetic fields are produced by charges is particularly untenable in quantum field theory, since it is essential for the understanding of such phenomena as the vacuum fluctuations of the electromagnetic field that the electromagnetic field have its own dynamical degrees of freedom, independently of the existence of charged matter.’’
\end{quote}
Wald presents this lesson from quantum field theory at the beginning of his book on classical electromagnetism, presumably because he thinks that we can learn about the proper formulation of classical electromagnetism by studying its successor, quantum electrodynamics.  We also think that debates within one theory can sometimes be informed by looking to deeper physics.  Ideally, these two theories should fit together neatly, with classical electromagnetism arising as a classical limit to quantum electrodynamics and quantum electrodynamics derivable by quantizing the classical electromagnetic field.

Having noted some reasons in favor of the above statistical strategy for explaining the arrow of electromagnetic radiation, let us now respond to four potential objections.  The first objection we will consider is the entirely reasonable request for more details.  In particular, a request for details on the correct probability distribution to apply over states of the electromagnetic field in the early universe---a request for details as to how the electromagnetic field should be incorporated into the Past Hypothesis and Statistical Postulate.  Unfortunately, those are not details that we can easily provide.  The microstates that one would be assigning probabilities over in the very early universe are not simply classical arrangements of charged matter and specifications of the state of the electromagnetic field.\footnote{There are a variety of problems that arise if you try to use such classical microstates for matter and field to develop a Boltzmannian statistical mechanics along the lines described in section \ref{PHsection}.  As is well known, classical attempts to explain the spectrum of black-body radiation failed and Max Plank derived the correct spectrum by appealing to quantum considerations \citep{Kuhn:1978aa}.  Here is a less well known problem: you cannot independently specify the states of matter and field to pick out a microstate. Gauss's law \eqref{gaussE} requires a certain coherence between states of matter and field.  However, states of the electromagnetic field that obey this constraint might still be unacceptable.  As \citet{hartensteinhubert} have shown, generic states of the electromagnetic field obeying the synchronic Maxwell equations, \eqref{gaussB} and \eqref{gaussE}, will give rise to pathological future behavior where shock fronts disrupt the dynamics and cause the theory to break down (see also \citealp[sec.\ 8.1]{lazarovici2018}).  One way to resolve the problems raised by \citet{hartensteinhubert} would be to adopt an approach where the electromagnetic field does not have any independent degrees of freedom, as in sections \ref{SRCsection}--\ref{WFsection}.}  At such a time, an adequate description of the physics would require quantum field theory (in particular, quantum electrodynamics). \citet{arntzenius1993}, \citet[pg.\ 1096]{north2003}, and \citet{atkinson2006} have discussed the importance of quantum electrodynamics for explaining the arrow of electromagnetic radiation.  We agree that the ultimate explanation of the arrow of radiation should appeal to quantum electrodynamics and understand that there is much work to be done.    Still, we think the simplified classical parable told in section \ref{FIELDsection} is helpful for getting a flavor for the kind of explanation that we expect quantum electrodynamics to yield.  It is correct in spirit, thought not in details.

A second objection to the above statistical explanation of the arrow of radiation is that it allows for backwards causation---deeming it merely improbable and not impossible.  We do not think the statistical explanation requires allowing for the possibility of backwards causation, though it has been paired with this view elsewhere.  \citet[pg.\ 1095]{north2003} writes:
\begin{quote}
``The temporally symmetric laws say that both advanced and retarded radiation could be emitted.  However, given the universe's thermal disequilibrium, the charges are overwhelmingly likely to radiate towards the future, as part of the overwhelmingly likely progression towards equilibrium in that temporal direction. They are overwhelmingly unlikely to radiate towards the past because the universe was at thermal equilibrium in that direction. Note that on this view the retarded nature of radiation is statistical: advanced radiation is not prohibited but given extremely low probability.''
\end{quote}
When North speaks of ``advanced radiation'' or ``radiat[ing] towards the past,'' she is talking about situations where there is a converging wave in the total electromagnetic field approaching a particular charge that resembles the advanced field of that charge (so that, locally, the advanced representation seems more natural than the retarded representation).  Such situations may be described as involving backwards causation.\footnote{Although we will generally view causes as preceding their effects, we see the appeal of allowing for causes that are in the future of their effects if there are periods of time (or regions of spacetime) where (relative to what we call past and future) entropy decreases and waves converge.  Boltzmann's hypothesis that the low-entropy of the early universe arose as a fluctuation from a high-entropy distant past would give rise to such periods of time before the early universe reached its low-entropy state \citep[ch.\ 10]{carroll2010}.}  But, they do not need to be.  Even when you consider a converging electromagnetic wave that can be represented by a purely advanced field (with no outgoing field), as in figure \ref{retarded-advanced}.b, you do not need to view the electromagnetic field at any point in space and time as caused by future charges.  You can instead view it as caused by earlier states of the electromagnetic field (see section \ref{EMsection}).  The wave moves towards the charge because it has been moving towards the charge.  In general, whether the electromagnetic field is purely retarded, purely advanced, or neither, it is possible to understand its time evolution purely in terms of forward causation.

A third objection that might be raised to our account is that the Past Hypothesis and Statistical Postulate, when spelled out precisely for electromagnetic field and matter, will be complicated.  The exact degree of complexity remains to be seen and the cost of that complexity will depend on whether one views these principles as laws of nature or as something else.  For now, let us just note that if you would like to adopt versions of the Past Hypothesis and Statistical Postulate to explain thermodynamic asymmetries, you cannot confine these principles to matter and ignore the electromagnetic field (seeking simplicity).  Specifying the initial state of the charged matter alone will fail to determine its future evolution because there are many states of the field compatible with any such state of charged matter (e.g., a given electromagnetic wave could be present or absent).  We need a way of selecting a particular state of the electromagnetic field, or of assigning probabilities over different states, if we want to be able to make predictions about the future motion of matter.

A fourth potential objection to our account is that we have not yet explained radiation reaction.  As was discussed in section \ref{EMsection}, for extended charged bodies radiation reaction can be explained by analyzing the way that electromagnetic waves propagate through such bodies on their way out.  The arrow of radiation reaction follows from the arrow of radiation.  That kind of explanation can go through even if the (retarded) waves emitted by the charged body are not the only electromagnetic waves in existence.  There may be other waves that were emitted by other charged bodies in the past or waves that are part of the free incoming electromagnetic field.  So long as those waves do not conspire to converge on the accelerating charge, we will see radiation reaction (as well as reaction to the forces from the other waves).  If all charges are extended charges, we can stop there.  For point charges, there are multiple ways of handling radiation reaction.  As was discussed in section \ref{EMsection}, one idea is to replace the Lorentz force law with the Lorentz-Dirac force law \eqref{lorentzdirac}.  If we treat the electromagnetic field at $t_0$ as a free incoming field, then the Lorentz-Dirac force law gives a well-defined equation of motion so long as $F_{inc}$ is well-defined at every point in spacetime that a charge passes through and no waves in the initial field converge precisely on any of the point charges.  Although our statistical explanation of the arrow of radiation does not deem such precisely converging waves impossible, they are effectively ruled out as they would only occur in a set of measure zero among the allowed initial conditions.  One should expect waves that converge on a region to be rare and waves that converge on a point to be absent.  Thus, the Lorentz-Dirac force law can be used to explain radiation reaction once statistical moves have been made to tame the free field.

\section{Strategy 2: The Sommerfeld Radiation Condition}\label{SRCsection}

An alternative strategy for explaining the arrow of radiation is to modify the laws of electromagnetism.  The cleanest way of doing this is by restricting the space of physical possibilities allowed by the theory to histories of matter and field where the electromagnetic field has no free (incoming) component in the retarded representation, $F=F_{ret}+F_{in}$:\footnote{Sommerfeld wrote down the original formulation of the condition in 1912 in order to have unique solutions to the Helmholtz equation \citep[for the history, see][]{Schot:1992aa}. This equation is time-independent, and the condition is accordingly a restriction on the solutions at \emph{spatial} infinity. This original boundary condition evolved into the above Sommerfeld Radiation Condition, requiring that all radiation be attributable to past sources. In his textbook, \citet[][p.\ 189]{Sommerfeld:1949aa} gives the following motivation for his boundary condition: ``We call it the \emph{condition of radiation}: the sources must be \emph{sources}, not \emph{sinks}, of energy. The energy which is radiated from the sources must scatter to infinity; \emph{no energy may be radiated from infinity into the prescribe singularities of the field} [\dots].''}
\begin{quote}
\textbf{The Sommerfeld Radiation Condition:}  The total electromagnetic field is purely retarded.  At every point in spacetime, $F_{in}=0$ and $F=F_{ret}$.
\end{quote}
This condition eliminates free fields from the retarded representation, though they would still be present if one chose to use the advanced representation, $F=F_{ret}=F_{adv}+F_{out}$.\footnote{See \citet[pg.\ 156--157]{frisch2005}.} If the electromagnetic field is purely retarded ($F_{in}=0$) at one time, it will be purely retarded at all times.  Thus it is equivalent to require that the field be purely retarded at one time, or to require, as above, that it be purely retarded at all times.

Assuming that there was an infinite past, the Sommerfeld Radiation Condition states that the electromagnetic field at a point in spacetime can be calculated by integrating contributions from progressively further distances and earlier times out to spatial and past infinity along the light-cone \eqref{retardedsolutions}.\footnote{See \citet[pg.\ 1087]{north2003}; \citet{price2006}; \citet[sec.\ 2.8]{earman2011}.}  If there was a first moment, one might attempt to impose a version of the Sommerfeld Radiation Condition by restricting the spatial integrals for the retarded potentials \eqref{retardedsolutions} so that the retarded times being integrated over never precede the first moment (as a way of positing that $F_{in}=0$ at the initial moment and thus at all future moments).  But, that will not work.  The recipe just described would have the consequence that, at the first moment, there is no electromagnetic field at any point in space where there is no charged matter---even right next to a charged body (in violation of Gauss's law, one of Maxwell's inviolable equations).  Instead, one might attempt to stipulate that the electromagnetic field at the first moment is just the field of each bit of charge at the first moment.  For example, a point charge at rest would be surrounded by a Coulomb electric field.  However, this strategy breaks down because the field generated by a bit of charged matter via \eqref{retardedsolutions} depends on its imagined past, and multiple fictional pasts will be compatible with the initial state of charged matter at the first moment \citep{hartensteinhubert}.  Thus, we do not see a precise way of stating the Sommerfeld Radiation Condition under the assumption of a first moment.\footnote{\citet[pg.\ 107]{frisch2005} suggests that we might ignore the Coulomb fields when applying the Sommerfeld Radiation Condition at a given moment such as the initial moment.  One way to do this, for point charges, would be to include, at the initial moment, only the generalized Coulomb field for each charge \citep[pg.\ 29]{zeh2007}.  This amounts to calculating the retarded fields that would have been generated if each particle had always been moving before the initial moment with the same velocity that they have at the initial moment.  \citet[sec.\ 3.3]{hartensteinhubert} show that there will be persistent and proliferating discontinuous jumps in the electromagnetic field values if you only match the velocities (and not the accelerations) between the hypothetical past trajectories and the actual future trajectories of charged particles. The fact that we do not observe such discontinuities speaks strongly against this proposal.}  To move forward with our assessment of this proposal, let us assume that there was an infinite past.

\subsection{Justification}

In his excellent and widely-used textbook on classical electromagnetism, \citet[pg.\ 446--447]{griffiths} gives the following justification for adopting the Sommerfeld Radiation Condition:
 \begin{quote}
``Although the advanced potentials are entirely consistent with Maxwell's equations, they violate the most sacred tenet in all of physics: the principle of \textbf{causality}.  They suggest that the potentials \emph{now} depend on what the charge and the current distribution \emph{will be} at some time in the future---the effect, in other words, precedes the cause.  Although the advanced potentials are of some theoretical interest, they have no direct physical significance.

``... the theory itself is \textbf{time-reversal invariant}, and does not distinguish `past' from `future.'  Time asymmetry is introduced when we select the retarded potentials in preference to the advanced ones, reflecting the (not unreasonable!) belief that electromagnetic influences propagate forward, not backward, in time.''
\end{quote}
Similar reasoning appears in \citet[pg.\ 346]{Schwinger:1998uq}; \citet{jefimenko2000};\footnote{Jefimenko incorporates the principle of causality into a broader vision regarding how fundamental laws in physics should be formulated:
\begin{quote}
``Causal relations between phenomena are governed by the \emph{principle of causality}.  According to this principle, all present phenomena are exclusively determined by past events.  Therefore equations depicting causal relations between physical phenomena must, in general, be equations where a present-time quantity (the effect) relates to one or more quantities (causes) that existed at some previous time.'' \citep[pg.\ 4]{jefimenko2000}
\end{quote}
Jefimenko's equations \eqref{jefimenko}, giving the current state of the electromagnetic field in terms of the past behavior of charges, fit this mold.} \citet{rohrlich2000,rohrlich2002,Rohrlich:2006aa,rohrlich}.  There are at least three points where one might criticize Griffiths' argument.  First, although it is true that, for the purely advanced potentials \eqref{advancedsolutions}, the value of the electromagnetic field at a given point in spacetime can be calculated mathematically by examining charges in the future (along the future light cone), it does not automatically follow that effects will precede their causes.  As we discussed in section \ref{EMsection}, advanced solutions can be interpreted with an ordinary order of cause and effect.  For example, the purely advanced converging wave in figure \ref{retarded-advanced}.b can be seen as a cause of the charge's motion (instead of as an effect that precedes this cause).  Second, to justify the use of purely retarded solutions Griffiths must reject not only purely advanced solutions, but also solutions that are neither purely retarded nor purely advanced and involve free fields whether they are expressed in the retarded or advanced representation. Finally, one could of course contest the ``principle of causality'' requiring causes to precede effects, though we will not explore that avenue here.

In a solution to Maxwell's equations that violates the Sommerfeld Radiation Condition and is not purely retarded, the value of the electromagnetic field at a given point in space cannot be fully attributed to past sources.  One could attempt to defend the Sommerfeld Radiation Condition by arguing that the electromagnetic field is created by charges and must always be fully attributable to past sources.  But, why make this assumption?  We see that defense as begging the question as to whether the condition should be adopted.

\citet[sec.\ 5]{frisch2000}\footnote{In his later book, \citet[pg.\ 152]{frisch2005} defends a variant of the Sommerfeld Radiation Condition that he calls the ``retardation condition'': ``\dots each charged particle physically contributes a fully retarded component to the total field.''  This causal claim appears to leave open the possibility of there being a genuinely free incoming field in addition to the retarded fields associated with charges.  Thus, we do not see in this claim any restriction on the space of physical histories allowed within classical electromagnetism.  (For Frisch, it is a claim about counterfactuals.)  Without restricting the allowed histories or assigning a probability distribution over them, we do not yet have the kind of resources that would be needed to explain the arrow of electromagnetic radiation.  \citet[pg.\ 152]{frisch2005} combines his retardation condition with a time-asymmetric assumption about the distribution of absorbing media that might be called an ``absorber condition'': ``\dots space-time regions in which we are interested generally have media acting as absorbers in their past. \dots fields that are not associated with charges that are relevant to a given phenomenon can generally be ignored, and it is easy to choose initial-value surfaces on which the incoming fields are zero.'' Setting the toothless retardation condition aside, this absorber condition could be used as part of a statistical explanation of the arrow of radiation (as it would require statistical reasoning to explain why certain media act as absorbers and not emitters---see \citealp[sec.\ 4]{frisch2000}; \citealp[sec.\ 3]{price2006}).  That being said, we do not think the condition is necessary to explain the asymmetry between converging and diverging waves.  A statistical explanation that assigns probabilities over states of the electromagnetic field in the early universe will render converging waves automatically improbable (section \ref{Ssection}).  There is no need to rely on assumptions about their eventual absorption.} argues that the Sommerfeld Radiation Condition is justified not by a deeper principle (like a principle of causality), but by the same kind of evidence that justifies Maxwell's equations.  We accept Maxwell's equations because of their success in explaining and predicting the behavior of charged matter and the electromagnetic field.  With the Sommerfeld Radiation Condition, the predictive and explanatory power of electromagnetic theory arguably increases, as this condition may be used to explain why electromagnetic waves generally diverge by ruling out certain solutions to Maxwell's equations containing converging waves.  If imposing the Sommerfeld Radiation Condition were the only way to explain the asymmetry of electromagnetic radiation, this would be a decisive argument for its inclusion among the laws of classical electromagnetism.  However, the presence of competing proposals leaves room for debate as to whether the condition should be adopted.

\subsection{Evaluation}\label{SRCEVALsection}

In favor of the Sommerfeld Radiation Condition, we believe that it is indeed capable of explaining why electromagnetic waves diverge, at least when combined with plausible statistical assumptions about the charged matter emitting the waves.  If there are no incoming free fields and each bit of charged matter makes a fully retarded contribution to the electromagnetic field, then it would require a carefully constructed arrangement of moving charges to form a wave that converges to a central point or region.  Such an arrangement would be exceedingly unlikely to occur in nature.  Although the Sommerfeld Radiation Condition can be used to explain the arrow of radiation, we do not think that the condition needs to be posited in order to explain the arrow.\footnote{\citet[pg.\ 507]{price2006} has argued that the Sommerfeld Radiation Condition is neither a necessary nor a sufficient condition for ``the observed asymmetry of radiation.''  Here is his argument that the Sommerfeld Radiation Condition is not sufficient:
\begin{quote}
``At least for some kinds of wave phenomena, there are possible solutions in which both $F_{in}=0$ and $F_{out}=0$. (Intuitively, imagine that the sources of ripples are surrounded by a good absorber, so that no waves can escape over the boundary $S$.) If such cases are possible---and mere possibility is enough, for the purposes of this point---then the [Sommerfeld Radiation Condition, $F_{in}=0$] cannot be sufficient for an observable asymmetry, on pain of contradiction. Whatever the observed asymmetry amounts to, it is certainly an asymmetry, and so could not consistently hold in both directions at once.'' \citep[pg.\ 502]{price2006}
\end{quote}
As an electromagnetic illustration of Price's description, imagine briefly oscillating a charge that emits an electromagnetic wave which is later absorbed by a spherical shell containing charges that are shaken by the wave, dissipating the wave's electromagnetic energy into heat (random motion).  In the time-reversed process, thermal motion of charges in the shell leads them to shake in unison and produce an electromagnetic wave that converges on the charge at the center.  The Sommerfeld Radiation Condition does not forbid this time-reversed process, but we have good reason to think it improbable.  Thus, even though the Sommerfeld Radiation Condition allows for the possibility of converging electromagnetic waves, it may still be used to explain why we do not observe them.  As Price puts the lesson of the quoted argument above: the Sommerfeld Radiation Condition does not \emph{characterize} the observed asymmetry of radiation that we seek to explain, but that leaves open the role that such a condition might play in \emph{explaining} the observed asymmetry.}  As we argued in the previous section, the arrow can be explained statistically without imposing such a condition.

In addition to explaining why electromagnetic waves diverge, the Sommerfeld Radiation Condition can also be used to explain the asymmetry of radiation reaction.  Standard textbook treatments of radiation reaction for extended charged bodies assume that radiation is fully retarded.\footnote{See \citet[ch.\ 16]{jackson}; \citet[sec.\ 11.2.3]{griffiths}; \citet[pg.\ 18]{frischpietsch2016}.}  However, as was discussed in section \ref{SEVALsection}, the radiation reaction force can be attributed to electromagnetic waves exiting an extended charged body without assuming that these sourced waves are the only electromagnetic waves present in nature.  If we shift our attention from extended charges to point charges, the Sommerfeld Radiation Condition alone will not be enough to derive radiation reaction forces (or to ensure conservation of energy and momentum).  That being said, one can combine the Sommerfeld Radiation Condition with some adjustment to the laws of electromagnetism to explain radiation reaction.  For example, one can adopt the Lorentz-Dirac force law \eqref{lorentzdirac}.  This law breaks down for certain free incoming fields, but it yields well-defined forces if there are no incoming fields---as would be ensured by the Sommerfeld Radiation Condition.

There are a number of criticisms that can be raised against using the Sommerfeld Radiation Condition to explain the arrow of electromagnetic radiation.  First, one might argue that there is an Occam's razor cost to adding another law to electromagnetism, complicating Maxwell's simple and elegant equations.  However, comparisons of simplicity are not so straightforward.  One can write new fundamental laws for electromagnetism giving the electric and magnetic fields directly in terms of the past behavior of charged matter via Jefimenko's equations,\footnote{See \citet[sec.\ 6.5]{jackson}; \citet{jefimenko2000}; \citet[sec.\ 10.2.2]{griffiths}.}
\begin{align}
\vec{E}(\vec{x},t)&=\int d^3\vec{x}' \left[ \frac{\rho(\vec{x}',t_r)}{|\vec{u}|^2}\hat{u}+\frac{\dot{\rho}(\vec{x}',t_r)}{c |\vec{u}|}\hat{u}-\frac{\dot{\vec{J}}(\vec{x}',t_r)}{c^2 |\vec{u}|}\right]
\nonumber
\\
\vec{B}(\vec{x},t)&=\frac{1}{c} \int d^3\vec{x}' \left[ \left(\frac{\vec{J}(\vec{x}',t_r)}{|\vec{u}|^2}+\frac{\dot{\vec{J}}(\vec{x}',t_r)}{c |\vec{u}|}\right) \times \hat{u}\right]
\ ,
\label{jefimenko}
\end{align}
using similar notation as in \eqref{retardedsolutions}.  Jefimenko's equations ensure that Maxwell's equations and the Sommerfeld Radiation Condition are all obeyed.  In our assessment, the original formulation of electromagnetism scores higher on simplicity and elegance as compared to the modified theory incorporating the Sommerfeld Radiation Condition.  But, there is room for disagreement.

Second, the rejection of incoming free fields puts charged matter and electromagnetic field in very different roles.  The field must be created by matter but matter need not be created by anything else.  As was discussed in section \ref{SEVALsection}, this clashes with more fundamental physics where charged matter and electromagnetic field are treated similarly.  There is no empirical evidence against the existence of incoming free fields (see section \ref{FIELDsection}), and we think it would be unnatural to deem them physically impossible.

Third, unlike the statistical approach, this strategy would not give a unified explanation as to why waves of all kinds diverge.  The Sommerfeld Radiation Condition can be used in an explanation as to why electromagnetic waves diverge, but it will be of little help in explaining why water or sound waves diverge.  One might argue that analogous assumptions about waves having sources can be used to explain asymmetries in other wave phenomena, but it will not be the Sommerfeld Radiation Condition itself that explains these asymmetries.  For example, the time-asymmetry of gravitational waves could potentially be explained by positing a radiation condition for gravity.\footnote{One may be inclined to subsume those different radiation conditions into a generalized radiation condition, which then appears to unify the different wave asymmetries. But this kind of radiation condition would still impose a different radiation condition for each domain. So this generalized radiation condition would not truly unify these asymmetries.}

Fourth, unlike the statistical approach, this strategy would not give a unified explanation of thermodynamic asymmetries and wave asymmetries.  The statistical account that we defended in the previous section gives a unified explanation of both using tools we already have and need (the Past Hypothesis and the Statistical Postulate).

As a fifth and final criticism, one might object that the Sommerfeld Radiation Condition can only explain the asymmetry between converging and diverging electromagnetic waves if one combines it with some kind of statistical story.  After all, even if the electromagnetic field is fully retarded, it is possible to have coordinated motions of charged bodies create converging electromagnetic waves.  One can argue that such motions are improbable by appealing to a probability distribution like the one generated by the Past Hypothesis and the Statistical Postulate.  However, if we take that probability distribution to range over states of both matter and field, then (as we argued in the previous section) there is no need to impose any additional conditions.  We already have a satisfactory explanation of the arrow of radiation.

\section{Strategy 3: Retarded Action-at-a-Distance}\label{RAADsection}

With the Sommerfeld Radiation Condition in place, the value of the electromagnetic field at the present location of a charge can always be traced back to the behavior of charges in the past.  In interactions between past and present charges, one might argue that the field is an unnecessary intermediary that can and should be eliminated (as Ritz argued in his debate with Einstein, \citealp{ritzeinstein1909}; \citealp[pg.\ 20]{frischpietsch2016}).  In this way, electromagnetism can be reformulated as a retarded action-at-a-distance theory where charges interact directly with one another across gaps in space and time  (\citealp{mundy1989}; \citealp{jefimenko2000}; \citealp{lange}; \citealp[sec.\ 2.5]{earman2011}; \citealp[pg.\ 460]{griffiths}).  For extended charged bodies, this can be done by calculating the Lorentz force density $\vec{f}=\rho \vec{E}+\frac{\rho}{c}\vec{v}\times\vec{B}$ using the retarded electric and magnetic fields that appear in Jefimenko's equations \eqref{jefimenko}, and then treating the result as a fundamental equation for electromagnetic force density in terms of the past behavior of charged matter.  For point charges, one can first take the limit as extended charges become point-size to derive the Li\'{e}nard-Wiechert potentials from the general expressions for the retarded potentials in \eqref{retardedsolutions}.  Then, one can use these potentials to generate point charge versions of Jefimenko's equations (\citealp[ch.\ 21]{feynman2}; \citealp[sec.\ 1.4]{jefimenko2000}; \citealp[sec.\ 10.3]{griffiths}).  From these, one can then use the Lorentz force law, $\vec{F}=q \vec{E}+\frac{q}{c}\vec{v}\times\vec{B}$, to eliminate the fields and construct a theory with unmediated retarded action-at-distance.  Each point charge acts as if it is responding to the retarded electric and magnetic fields produced by every other point charge, though these fields are not part of the theory's fundamental ontology---its base-level account as to what exists.  (These fields can be treated either as useful fictions or as non-fundamental but still real.)  Because charges respond to the (fictional/non-fundamental) fields of other charges but not to their own fields, we avoid the problem for point charges of ill-defined self-interaction from section \ref{EMsection}.

Because this strategy for explaining the arrow of radiation is so similar to the last, we can move directly into an assessment of its strengths and weaknesses.

\subsection{Evaluation}\label{RAADevaluation}

On this approach, the electromagnetic field is treated as a non-fundamental entity or as a calculational tool, not as part of the fundamental ontology of the theory.  That being said, the values that the field takes are exactly the same as in the previous proposal where the Sommerfeld Radiation Condition is imposed to eliminate any incoming free field.  Thus, just as the Sommerfeld Radiation Condition can explain why converging electromagnetic waves are rare, so can a retarded action-at-distance formulation of electromagnetism.  As before, converging waves would require precise arrangements of accelerating charges that can reasonably be treated as improbable.

In comparison to a formulation of electromagnetism that includes the Sommerfeld Radiation Condition, a retarded action-at-distance formulation has a sparser ontology and, one might argue, simpler laws.  At the bottom level, there are only charged bodies.  There is no electromagnetic field.  Maxwell's equations, the Lorentz force law, and the Sommerfeld Radiation Condition are not fundamental laws.  In their place, we have a single (rather complicated) force law.  For point charges, the force on a point charge $q_k$ with velocity $\vec{v}$ is
\begin{align}
\vec{F}_k&=\sum_{j\neq k} - q_k q_j\left\{\left(\frac{\hat{u}_j}{|\vec{u}_j|^2}+\frac{\vec{u}_j}{c}\frac{d}{dt}\left(\frac{\hat{u}_j}{|\vec{u}_j|^2}\right)+\frac{1}{c^2}\frac{d^2 \hat{u}_j}{dt^2} \right)\right.
\nonumber
\\
&\qquad\left.+\frac{\vec{v}}{c}\times \left[ \hat{u}_j \times \left(\frac{\hat{u}_j}{|\vec{u}_j|^2}+\frac{\vec{u}_j}{c}\frac{d}{dt}\left(\frac{\hat{u}_j}{|\vec{u}_j|^2}\right)+\frac{1}{c^2}\frac{d^2 \hat{u}_j}{dt^2}\right) \right]\right\}
\ ,
\label{oneforcelaw}
\end{align}
where $j$ is an index on the particles, $\vec{u}_j=\vec{x}-\vec{x}_j(t_{r,j})$ is the vector pointing to $q$ now from the position of the $j$-th charge at the retarded time $t_{r,j}$ (when the backwards light cone of $q$ intersects the world line of the $j$-th charge), and $\hat{u}_j$ is the unit vector pointing in the same direction.\footnote{To arrive at \eqref{oneforcelaw}, one can plug the electric and magnetic fields for a point charge from \citet[sec.\ 21-1]{feynman2} (rewritten in cgs units) into the Lorentz force law.}  As compared to \eqref{oneforcelaw}, we find Maxwell's equations and the the Lorentz force law to be a more attractive set of laws.  But, there is a certain appeal to having a single law.

In addition to explaining the arrow of electromagnetic radiation, one can also attempt to explain radiation reaction within a retarded action-at-a-distance formulation of electromagnetism.  For extended charges, the explanation will be just as in the previous section.  One can derive the radiation reaction forces by considering (retarded) interactions between different pieces of a charged body.  For point charges, the force law given above \eqref{oneforcelaw} cleanly avoids infinite or ill-defined self-interaction but would not yield any radiation reaction force, because it is derived by only including ordinary Lorentz forces from the retarded fields of other charges.  However, one might attempt to incorporate radiation reaction by, for example, starting instead from the Lorentz-Dirac force law and then eliminating the electromagnetic fields that appear in it.

In section \ref{SRCEVALsection}, we criticized the Sommerfeld Radiation Condition for treating charged matter and the electromagnetic field as fundamentally different sorts of things (whereas, in our opinion, quantum field theory points to them being the same sort of thing).  A retarded action-at-a-distance version of electromagnetism goes even further, not just viewing the electromagnetic field as a mere emanation from charged matter but eliminating it entirely.

As with the Sommerfeld Radiation Condition, the explanation as to why electromagnetic waves diverge given by a retarded action-at-a-distance version of electromagnetism would be very different from the explanations as to why non-electromagnetic waves (like sound waves and water waves) diverge.  Also, the explanation for electromagnetic wave asymmetries would be distinct from explanations of thermodynamic asymmetries.

Again repeating a criticism of the Sommerfeld Radiation Condition, one would need to appeal to statistical assumptions regarding the motions of charges within a retarded action-at-a-distance formulation of electromagnetism to explain the arrow of electromagnetic radiation.  Once statistical reasoning is brought in, why not give a fully statistical explanation of the arrow (as in section \ref{Ssection})?

Beyond these disadvantages inherited from the Sommerfeld Radiation Condition approach, the retarded action-at-a-distance strategy comes with a couple problems of its own.  First, energy and momentum will not be conserved (\citealp{mundy1989}; \citealp[ch.\ 5]{lange}; \citealp[pg.\ 496]{earman2011}; \citealp{potentialenergy}).  The ordinary derivations of energy and momentum conservation in classical electromagnetism assume that the electromagnetic field carries both energy and momentum \citep[ch.\ 8]{griffiths}.\footnote{For extended distributions of charge, conservation of energy and momentum is straightforward.  For point charges, things are not so straightforward because radiation reaction complicates derivations of energy and momentum conservation \citep[sec.\ 4.2]{lazarovici2018}.}  If the electromagnetic field is eliminated, then the field's energy and momentum are absent and the conservation laws no longer hold.  For example, if two positively charged bodies are shot straight at each other, they will repel one another, initially slowing down and losing kinetic energy.  Normally, we would say that during this process energy is transferred from the bodies to the electromagnetic field.  But, if the field does not exist then the energy is simply lost.  One might try to recover conservation of energy by relocating the energy of the field in the charged bodies and viewing it as potential energy determined by the current arrangement of charged bodies.  But, this will not always be possible because the energy in the field depends on the past motions of charged matter.

The second problem that is unique to the retarded action-at-a-distance strategy is non-locality: because the electromagnetic field does not exist, interactions between charged bodies are spatially and temporally non-local.  This is a feature that the approach wears on its sleeve in the name ``action-at-a-distance.''  It is not obvious why or whether we should desire spatially and temporally local theories \citep[sec.\ 4.3]{lazarovici2018}.  \citet{lange} presents an in-depth analysis of locality in classical electromagnetism, arguing that the best version of the theory includes fields and is both spatially and temporally local.  But, he does not argue that this version is best because it is local.  Instead, he argues for it by other means (building on the problems with energy and momentum conservation).  We agree that non-locality is not the main problem for retarded action-at-a-distance.  There are plenty of other reasons to dislike the approach.

\section{Strategy 4: The Wheeler-Feynman Theory}\label{WFsection}

Before we get to the Wheeler-Feynman half-retarded half-advanced action-at-a-distance theory,\footnote{The Wheeler-Feynman theory and its explanation of the arrow of radiation are discussed in \citet[sec.\ 5.7 and 5.8]{davies}; \citet[sec.\ 5]{arntzenius1993}, \citet[ch.\ 6]{frisch2005}; \citet{price1996, price2006, lazarovici2018}.} let us first consider a variant that includes the electromagnetic field.  In section \ref{SRCsection}, we considered supplementing classical electromagnetism with the Sommerfeld Radiation Condition, requiring that in the retarded representation \eqref{Rrepresentation} there is no (incoming) free field and thus that the total electromagnetic field is fully retarded.  Adopting the half-retarded half-advanced representation---\eqref{Grepresentation} with $\alpha=1/2$---we can impose a similar condition to rule out any (half-incoming half-outgoing) free field and ensure that the electromagnetic field is half-retarded half-advanced:
\begin{quote}
\textbf{The Wheeler-Feynman Radiation Condition:}  The total electromagnetic field is half-retarded half-advanced.  At every point in spacetime, $\frac{1}{2}F_{in}+\frac{1}{2}F_{out}=0$ and $F=\frac{1}{2}F_{ret}+\frac{1}{2}F_{adv}$.
\end{quote}
Now, we can remove the electromagnetic field to get an action-at-a-distance theory---as in the transition from a field theory in section \ref{SRCsection} to  a retarded action-at-a-distance theory in section \ref{RAADsection}.  The primary attraction of the Wheeler-Feynman action-at-a-distance theory is that it enables a derivation of the Lorentz-Dirac equation of motion for point particles \eqref{lorentzdirac}, which includes radiation reaction and excludes any infinite or ill-defined self-interaction.  Thus, we will assume in this section that we are dealing with point charges and not charge distributions.\footnote{\citet{Bauer:2013aa} discuss the Wheeler-Feynman theory for charge distributions.}

\begin{figure}[htb]\centering
\includegraphics[width=8 cm]{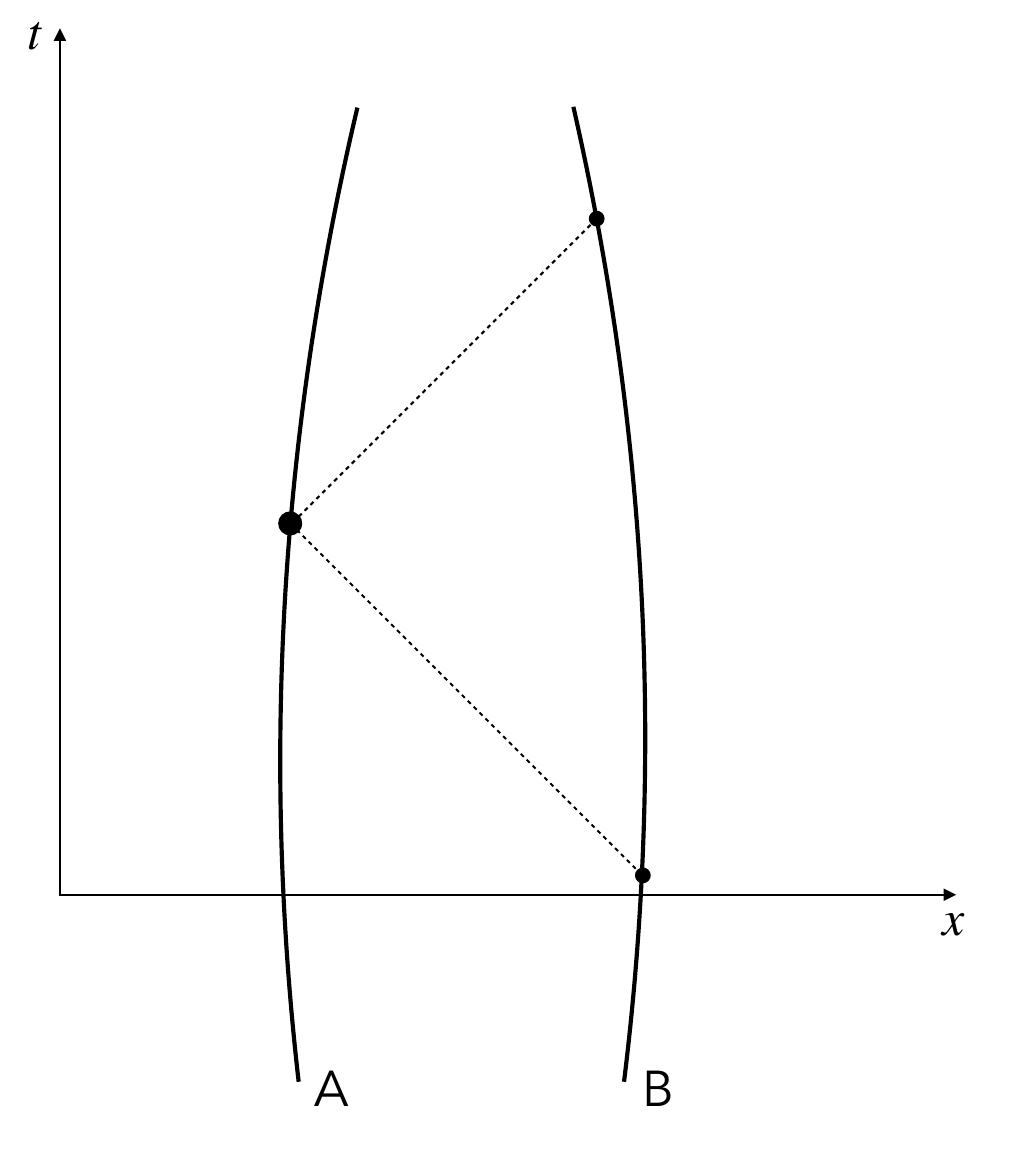}
\caption{The Wheeler-Feynman theory is an action-at-a-distance theory where interactions are half-retarded and half-advanced.  The force that charge $B$ exerts on charge $A$ at the present moment is the sum of a retarded force from $B$'s earlier state (when $B$ crosses the past light-cone of $A$) and an advanced force from $B$'s later state (when $B$ crosses the future light-cone of $A$).}
\label{fig:direct-interaction}
\end{figure}

In the Wheeler-Feynman action-at-a-distance theory, the force on point charge $k$ at $(\vec{x}_k,t)$ from every other charge $j$ is\footnote{This force law has been called the Fokker-Tetrode-Schwarzschild equation \citep{Bauer:2013aa} and can be written concisely in relativistic notation as
\begin{equation}
\label{eq:law-wf-2}
m_{k}\ddot{z}^{\mu}_{k}(\tau)=\frac{q_{k}}{c}\sum_{j\neq k} ^{N}  \frac{1}{2} \left(F^{\mu\nu}_{(j)ret}(z_{k}(\tau))+F^{\mu\nu}_{(j)adv}(z_{k}(\tau))\right)\dot{z}_{k\nu}(\tau), 
\end{equation}
where $z_{k}(\tau)$ is the trajectory of the $k$th charge in Minkowski space-time, $\tau$ the corresponding proper time, and a dot indicates derivation with respect to proper time.
} 
\begin{align}
\label{eq:law-wf}
\vec{F}_k(\vec{x}_k,t)=q_{k}\sum_{j\neq k} ^{N} \left(\frac{1}{2}\left(\vec{E}_{ret}^{(j)}(\vec{x}_k,t)+\vec{E}_{adv}^{(j)}(\vec{x}_k,t)\right)
  +\frac{\vec{v}_k}{c}\times\frac{1}{2}\left(\vec{B}_{ret}^{(j)}(\vec{x}_k,t)+\vec{B}_{adv}^{(j)}(\vec{x}_k,t)\right)\right)
  \ ,
\end{align}
where we have simplified the mathematics by helping ourselves to the retarded and advanced electric and magnetic fields of each charge, even though these fields are not part of the theory's fundamental ontology (the fields are either useful fictions or non-fundamental entities).\footnote{The retarded and advanced electric and magnetic fields in \eqref{eq:law-wf} are calculated from the retarded and advanced Li\'{e}nard--Wiechert potentials of each particle via equation \eqref{fieldsfrompotentials}.}  In reality, the force from particle $j$ on particle $k$ is the combination of a direct interaction with particle $j$ in the past---at the point in spacetime where its trajectory intersects the past light-cone of $(\vec{x}_k,t)$---and a direct interaction with particle $j$ in the future---at the point in spacetime where its trajectory intersects the future light-cone of $(\vec{x}_k,t)$ (see figure \ref{fig:direct-interaction}).  The first interaction is naturally interpreted as forwards causation and the second as backwards causation.  In \eqref{eq:law-wf}, self-interactions are excluded from the outset.  The force on charge $k$ comes solely from the other charges, $ j\neq k$.  Wheeler and Feynman sought to derive the correct force of radiation reaction by viewing it as the result interaction with other charges, not self-interaction.  Let us now turn to their derivation.

\subsection{Time-Asymmetry}

Thus far, no time-asymmetry has been introduced.  The Wheeler-Feynman theory is time-symmetric.  Somehow, we need to find explanations for both the asymmetry of electromagnetic radiation (why waves diverge) and the asymmetry of radiation reaction.  Let us first see how Wheeler and Feynman sought to explain radiation reaction as the result of half-retarded half-advanced action-at-a-distance between the accelerating charge and the charges that surround it.\footnote{Without imposing assumptions on the arrangement of charges surrounding the accelerating charge, you cannot derive the correct radiation reaction force.  For example, if the universe contains only a single charged particle then the Wheeler-Feynman theory predicts that there will be no radiation reaction force.}  We will review the most general derivation given by \citet[][pg.\ 169--171]{Wheeler:1945aa}, where they postulate a condition on the absorbing behavior of the surrounding particles to derive the Lorentz-Dirac equation (\citealp[see also][Ch.\ 6]{frisch2005}; \citealp[][]{Forgione:2020aa}).  In the end, we will see that Wheeler and Feynman's strategy for explaining radiation reaction also yields an explanation as to why electromagnetic waves diverge.

\begin{figure}[htb]\centering
\includegraphics[width=8 cm]{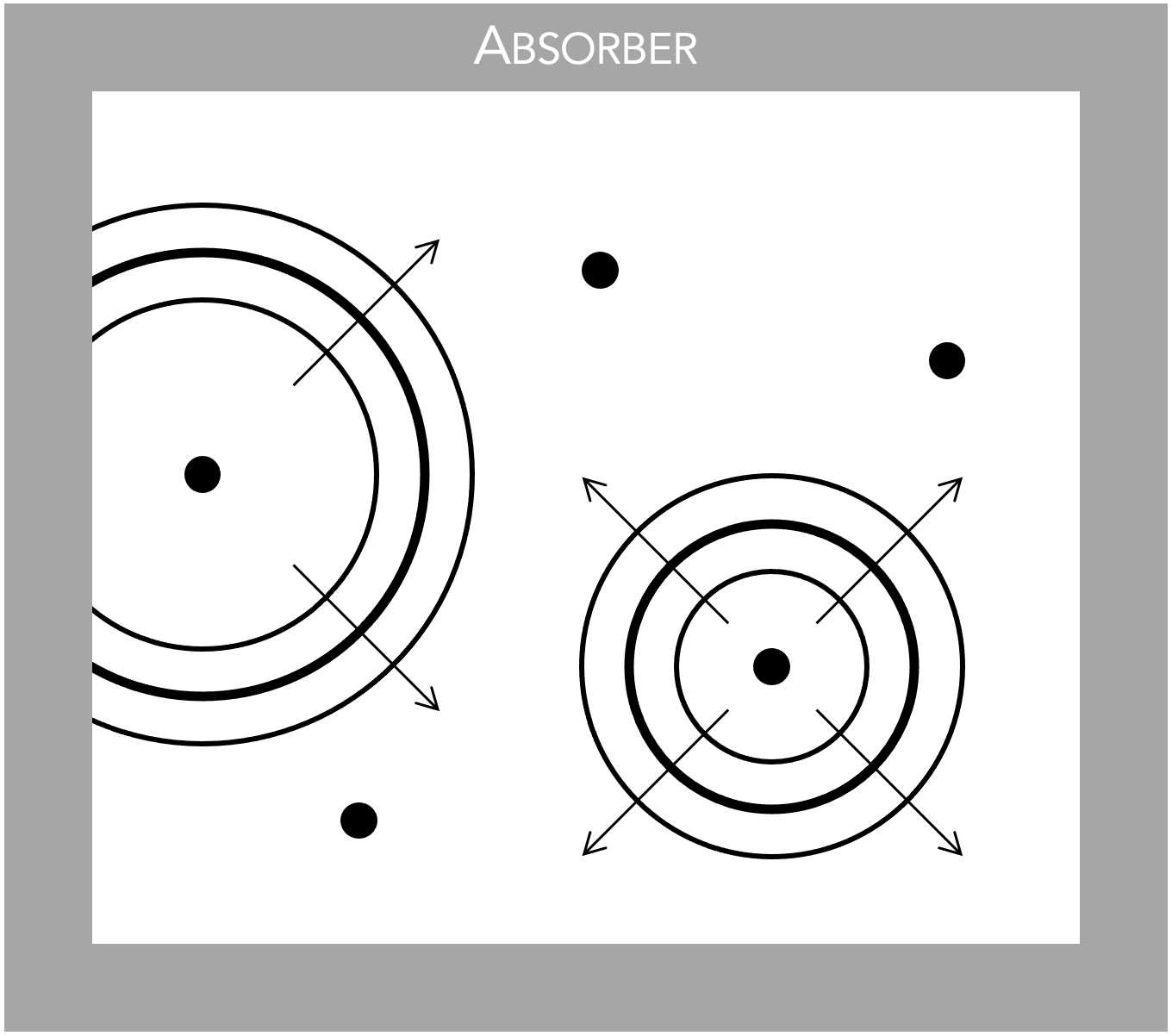}
\caption{Wheeler and Feynman's Absorber Condition posits the existence of an absorber surrounding the charges under consideration, stopping any electromagnetic waves from escaping the region.  The absorber is shown here as nearby, but in reality it could be very far away.}
\label{fig:absorber}
\end{figure}

Wheeler and Feynman's absorber condition (figure \ref{fig:absorber}) requires that the charges whose motion we are studying be surrounded by an absorber that completely absorbs any radiation they emit, where ``complete absorption means that a test charge placed anywhere outside the absorbing medium will experience no disturbance'' \citep[][pg.\ 169]{Wheeler:1945aa}.  It is hard to see how such perfect absorption could occur or why we should assume that it does, but let us press on to see what can be achieved by adopting this idealization.  In precise terms, Wheeler and Feynman's assumption is the following:
\begin{quote}
\textbf{The Absorber Condition:}  All of the charges under consideration are surrounded by an absorber such that anywhere outside the absorber the total field (which is the sum of the half-retarded half-advanced fields of the charge under consideration and all of the other charges, including those in the absorber) is zero,
\begin{equation}
\label{eq:absorber-condition}
\sum_{j}\left(\frac{1}{2}F_{ret}^{(j)}+\frac{1}{2}F_{adv}^{(j)}\right) = 0  \quad \text{(outside the absorber)}
\ .
\end{equation}
\end{quote}
From the absorber condition, it follows that the individual sums in \eqref{eq:absorber-condition} are zero outside the absorber: $\sum_{j}\frac{1}{2}F_{ret}^{(j)}=0$ and $\sum_{j}\frac{1}{2}F_{adv}^{(j)}=0$.  That is because, outside the absorber, the first sum represents diverging waves and the second sum represents converging waves.  It is impossible for these waves to fully destructively interfere if the sums are not individually zero. Because the sums are individually zero, their difference is zero:
\begin{equation}
 \sum_{j}\left(\frac{1}{2}F_{ret}^{(j)}-\frac{1}{2}F_{adv}^{(j)}\right)=0  \quad \text{(outside the absorber)}
 \ .
\end{equation}
This difference is a solution to the free Maxwell equations (because of the minus-sign).  If it is always zero outside the absorber, it must be zero inside as well,
\begin{equation}
\label{eq:absorber-everywhere}
\sum_{j}\left(\frac{1}{2}F_{ret}^{(j)}-\frac{1}{2}F_{adv}^{(j)}\right)=0  \quad \text{(everywhere)}.
\end{equation}

To calculate the force on a given particle $k$, we need to analyze the sum of the half-retarded half-advanced fields of all other particles at the location of $k$,
\begin{equation}
\label{eq:wf-interaction}
\sum_{j\neq k}\left(\frac{1}{2}F_{ret}^{(j)}+\frac{1}{2}F_{adv}^{(j)}\right)
\ .
\end{equation}
This can be rewritten as 
\begin{equation}
\label{eq:decomposition-wf-ret}
\sum_{j\neq k}F_{ret}^{(j)}+\left(\frac{1}{2}F_{ret}^{(k)}-\frac{1}{2}F_{adv}^{(k)}\right)-\underbrace{\sum_{\text{all}\,j}\left(\frac{1}{2}F_{ret}^{(j)}-\frac{1}{2}F_{adv}^{(j)}\right)}_\text{$=0$}
\ ,
\end{equation}
where \eqref{eq:absorber-everywhere} has been used to eliminate the final term.  The first term is the sum of the fully retarded fields associated with every other particle, which would give us a retarded action-at-a-distance theory (as in section \ref{RAADsection}) if it were the only term.  This is the first term in the Lorentz-Dirac equation.  Although it is not immediately apparent, the second term in \eqref{eq:decomposition-wf-ret} captures radiation reaction.  It can be replaced by a formula due to Dirac, yielding the second term in the Lorentz-Dirac equation \eqref{lorentzdirac}.  Thus, we have arrived at the Lorentz-Dirac equation (without any free incoming field) and a potential explanation as to the time-directed nature of radiation reaction.  Also, because the non-radiation-reaction force on a charge can be attributed to the retarded fields of other charges, we appear poised to give an explanation as to why electromagnetic waves diverge analogous to the one in section \ref{RAADsection} based on retarded action-at-a-distance.

Pausing here, it may look like we have derived time-asymmetric consequences from time-symmetric assumptions.  In fact, \eqref{eq:decomposition-wf-ret} is not alone sufficient to explain either the arrow of radiation reaction or the arrow of radiation.  To see the problem, note---as \citet[][pg.\ 170]{Wheeler:1945aa} do---that we can also derive the following equation for the field acting on charge $k$ \eqref{eq:wf-interaction},
\begin{equation}
\label{eq:decomposition-wf-adv}
\sum_{j\neq k}F_{adv}^{(j)}-\left(\frac{1}{2}F_{ret}^{(k)}-\frac{1}{2}F_{adv}^{(k)}\right)+\underbrace{\sum_{\text{all}\,j}\left(\frac{1}{2}F_{ret}^{(j)}-\frac{1}{2}F_{adv}^{(j)}\right)}_\text{$=0$}
\ ,
\end{equation}
in the same way that \eqref{eq:decomposition-wf-ret} was derived.  Using \eqref{eq:decomposition-wf-adv} to derive the force on $k$, the first term yields a non-radiation-reaction force that can be attributed to the fully advanced fields of the other particles and the second term yields a radiation reaction force that is opposite in sign to the force derived from the second term of \eqref{eq:decomposition-wf-ret}.  The Absorber Condition ensures that the force on $k$ can be calculated equivalently from either \eqref{eq:decomposition-wf-ret} or \eqref{eq:decomposition-wf-adv}.  The Absorber Condition allows for both converging and diverging electromagnetic waves in the total half-retarded half-advanced electromagnetic field.

To get time asymmetry, \citet[][pg.\ 170]{Wheeler:1945aa} assume that ``In our example [of an accelerated charge surrounded by an absorber] the particles of the absorber were either at rest or in random motion before the time at which the impulse was given to the source. \dots the sum \dots of the retarded fields of the adsorber [sic] particles had no particular effect on the acceleration of the source.''  If we number the particles surrounded by the absorber $1$ through $N$ and the particles of the absorber $N+1$ through $M$, \eqref{eq:decomposition-wf-ret} can be rewritten as
\begin{equation}
\sum_{j\neq k}^NF_{ret}^{(j)}+\sum_{j=N+1}^MF_{ret}^{(j)}+\left(\frac{1}{2}F_{ret}^{(k)}-\frac{1}{2}F_{adv}^{(k)}\right)
\ .
\end{equation}
Wheeler and Feynman are claiming that the second sum is (at least approximately) zero.  Let us label this as another condition:
\begin{quote}
\textbf{The Second Absorber Condition:}  The sum total of the retarded fields of the particles in the absorber ($N+1$ through $M$) at the location of any particle under consideration is negligible:
\begin{equation}
\sum_{j=N+1}^MF_{ret}^{(j)}\approx 0
\ .
\end{equation}
\end{quote}
One can ask why we should assume that the Second Absorber Condition holds.  \citet[][pg.\ 170]{Wheeler:1945aa} appeal to statistical considerations: ``We have to conclude with Einstein that the irreversibility of the emission process is a phenomenon of statistical mechanics connected with the asymmetry of the initial conditions with respect to time.''  To explain their dismissal of a converging wave solution, \citet[][pg.\ 170]{Wheeler:1945aa} write ``No electrodynamic objection can be raised against this solution of the equations of motion. Small \emph{a priori} probability of the given initial conditions provides the only basis on which to exclude such phenomena.''

\begin{figure}[htb]\centering
\includegraphics[width=11 cm]{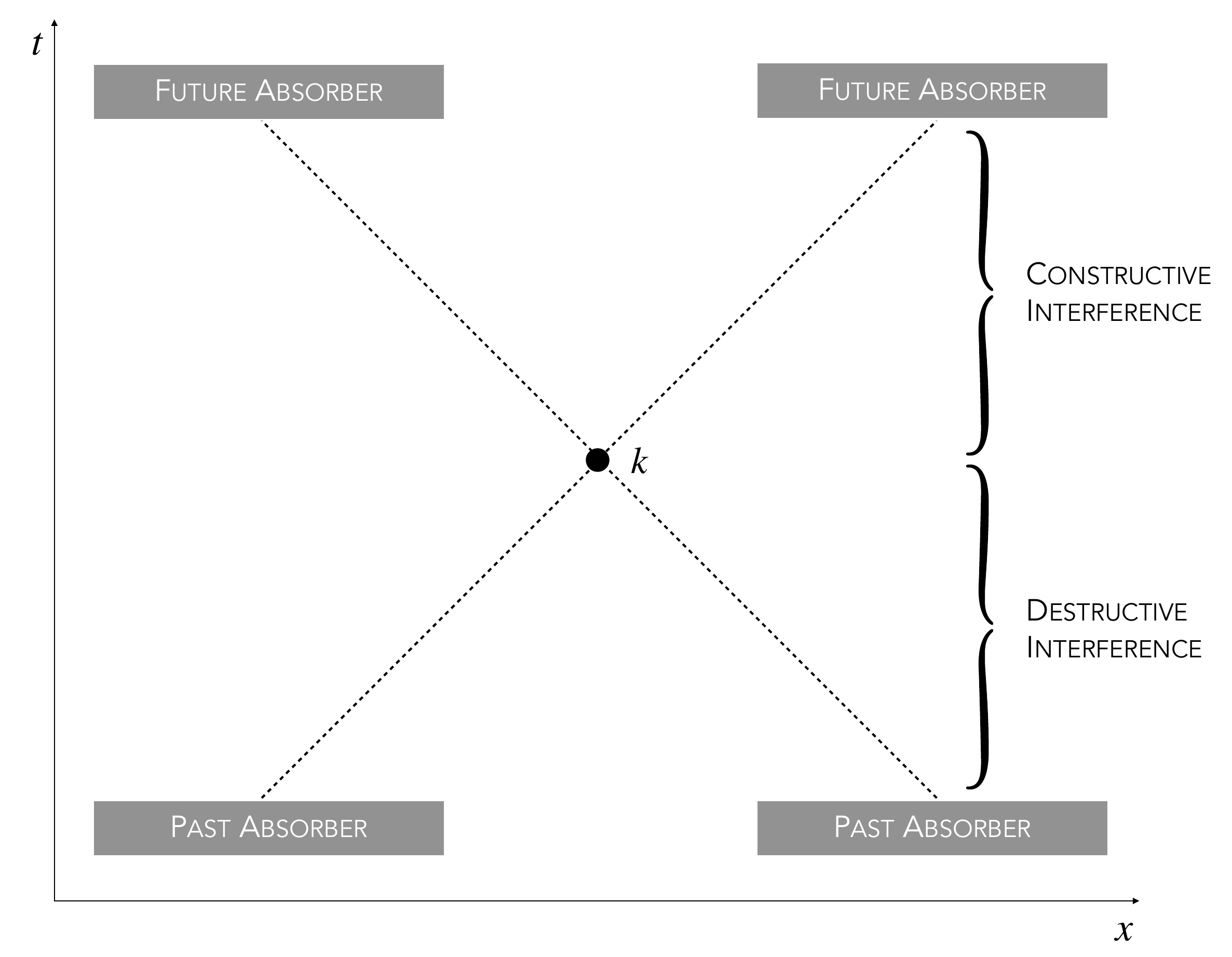}
\caption{In the Wheeler-Feynman theory with the two absorber conditions, if a charge is briefly shaken then after the shaking there will be constructive interference between the half-retarded field of the charge and the half-advanced fields of the other charges.  This yields the usual diverging wave (figure \ref{retarded-advanced}.a).  Before the shaking, there will be destructive interference between the half-advanced field of the charge and the half-advanced fields of the other charges.  Thus, the shaking will not generate a contribution to the total field in the past.}
\label{fig:interference}
\end{figure}

The Second Absorber Condition requires that the past behavior of the absorber particles is sufficiently disordered that the retarded fields do not combine to form converging waves.  By contrast, the future behavior of absorber particles is sufficiently ordered that the advanced waves do combine to form diverging waves.  With the two absorber conditions in place, if you briefly shake a single charge (as in figure \ref{retarded-advanced}.a) the total half-retarded half-advanced field after shaking will be the sum of a diverging retarded wave from the charge, of half the usual strength, and an equally strong diverging advanced wave from the other charges (figure \ref{fig:interference}).  Although the total field is not part of the theory's fundamental ontology, we can recognize that other particles within the absorber will react as if they are experiencing a force from this diverging wave.

\subsection{Evaluation}

In evaluating the Wheeler-Feynman half-retarded half-advanced action-at-a-distance theory, the pros and cons closely resemble those for a retarded action-at-a-distance version of electromagnetism (section \ref{RAADsection}).  For point charges, either kind of action-at-a-distance theory will easily avoid infinite or ill-defined self-interaction.  The main difference is the handling of radiation reaction.  To the extent that the two absorber conditions are plausible, the Wheeler-Feynman theory has a way of deriving radiation reaction from interaction with other charges.  By contrast, a simple retarded action-at-a-distance version of electromagnetism beginning with the Lorentz force law will fail to incorporate radiation reaction.  Still, a more complicated version that begins with the Lorentz-Dirac force law could include radiation reaction (see section \ref{RAADevaluation}).

As was the case for retarded action-at-a-distance, it will be possible to explain the arrow of electromagnetic radiation in Wheeler-Feynman electrodynamics by assigning low probabilities to histories where particles generate converging waves.  The exact details as to how one assigns probabilities will be complicated because one cannot simply assign probabilities to states at a time and ask how those states evolve (as was discussed in section \ref{PHsection}).  In the Wheeler-Feynman theory, the time evolution depends on both the past and the future.  Although a statistical explanation of the arrow of radiation should be possible, its complexity may make it unattractive.

The Second Absorber Condition can potentially be given a statistical explanation once the first is in place, but why should we assume that charged matter around us is always surrounded by a perfect absorber?  In other words, why should we accept the original Absorber Condition?  The condition is not merely a constraint on the state of the universe at a time, but rather a global constraint requiring distant charges to behave in a very specific manner.  Supporters of the Wheeler-Feynman theory have proposed relaxing or removing the original Absorber Condition \citep{Hogarth:1962uq, Hoyle:1995aa, Bauer:2014wq,lazarovici2018}, arguing that there are other ways to explain the arrows of radiation and/or radiation reaction within the Wheeler-Feynman theory.  We will not explore such proposals here.

We have repeatedly objected to approaches that treat the electromagnetic field and charged matter as very different kinds of things (sections \ref{SEVALsection}, \ref{SRCEVALsection}, and \ref{RAADevaluation}).  Wheeler-Feynman electrodynamics eliminates the electromagnetic field from the fundamental ontology and thus does not treat it in a similar manner to charged matter.  As \citet[][pg.\ 426]{Wheeler:1949aa} put it, ``There is no such concept as `the' field, an independent entity with degrees of freedom of its own.''  Our objection to such differing treatment of matter and field was that it does not fit well with quantum field theory, where the electromagnetic field looks very similar to charged matter. Feynman had hoped that the Wheeler-Feynman approach would lead to a better version of quantum field theory \citep{feynman1965}, but that dream has not yet been realized.

In the Wheeler-Feynman theory, the explanation as to why electromagnetic waves diverge is distinct from the explanations as to why other waves diverge.  The half-retarded half-advanced nature of electromagnetic waves is not a general feature of wave phenomena.  The  Wheeler-Feynman explanation as to why electromagnetic waves diverge is somewhat akin to the explanations of thermodynamic asymmetries (in that the explanations are all statistical), but it is not clear how these explanations are going to be bundled together because one cannot simply assign a probability distribution over initial conditions.  The combined account will not be based on a past hypothesis and statistical postulate of the type discussed in section \ref{Ssection}.

The Wheeler-Feynman theory involves interactions that occur across spatiotemporal gaps both forward and backward in time.  It is thus a non-local theory with backwards causation.  These are oddities that can be stomached, but it is important to recognize that there are other versions of electromagnetism that avoid them.

The elimination of the electromagnetic field in Wheeler-Feynman electrodynamics means that the theory violates conservation of energy and momentum, at least if we assume that those quantities are features of the state of the universe at a time.  There have been attempts to recover a kind of conservation by allowing these quantities to be determined by the past and the future as well as the present (\citealp{Wheeler:1949aa}; \citealp[pg.\ 155]{lazarovici2018}).  Although this kind of move may be acceptable for justifying the use of energy and momentum conservation in calculations, we do not find it appealing as a description of energy and momentum in nature.

Lastly, we want to point out the main mathematical difficulty in the Wheeler-Feynman theory. Because the electromagnetic field is removed and interactions are half-retarded and half-advanced, the dynamical equations are not standard differential equations but delay-differential equations. They relate the force on a charge at one time to the states of other particles at different times. These laws do not (in general) allow one to formulate and solve initial value problems, as in other physical theories, by specifying the complete physical state at a time and evolving the state forwards. Therefore, solving the Wheeler-Feynman equations is cumbersome (even for simple systems) and very few precise results have been calculated \citep[see][for an overview on how to tackle delay-differential equations]{hartensteinhubert}.

\section{Conclusion}

The arrow of radiation is one of many arrows of time that appear in physics. We have defended the view that the arrow of radiation can be absorbed into a larger statistical framework that also explains the other arrows. We regard this unification as one of the most compelling advantages of the account.  The time-asymmetric behavior of electromagnetic waves is explained in the same way as the time-asymmetric behavior of water waves, sound waves, melting ice cubes, deflating tires, and so much more. Another compelling advantage of this strategy for explaining the arrow of electromagnetic radiation is that we do not need to change the laws of electromagnetism, as is done in each of the competing strategies.  Also, matter and field are treated as similar kinds of things, each possessing independent degrees of freedom.

The point of disagreement between our preferred strategy and the others is not whether the explanation of the arrow of radiation should be statistical or not, but whether it should be purely statistical or only partially statistical.  As we have seen, strategies 2 through 4 (that is, the Sommerfeld Radiation Condition, the retarded action-at-a-distance theory, and the Wheeler--Feynman theory) will ultimately need to invoke some statistical reasoning to exclude converging waves.  If statistical assumptions are needed on any account, why not embrace a fully statistical account of the arrow of radiation?

\section*{Acknowledgements}
We wish to thank Craig Callender, Eddy Keming Chen, Erik Curiel, Olivier Darrigol, Sheldon Goldstein, Tim Maudlin, John Norton, Jill North, Sterl Phinney, Jos Uffink, David Wallace, and the anonymous reviewers for helpful comments and discussions.  Open access publication has been funded by a Faculty Support Grant of The American University in Cairo.


\begin{thebibliography}{}

\bibitem[\protect\citename{Albert, }2000]{albert2000}
Albert, David~Z. 2000.
\newblock {\em Time and Chance}.
\newblock Harvard University Press.

\bibitem[\protect\citename{Albert, }2015]{Albert:2015aa}
Albert, David~Z. 2015.
\newblock {\em After Physics}.
\newblock Cambridge, MA: Harvard University Press.

\bibitem[\protect\citename{Allori, }2015]{allori2015}
Allori, Valia. 2015.
\newblock Maxwell's Paradox: The Metaphysics of Classical Electrodynamics and
  its Time-Reversal Invariance.
\newblock {\em Analytica}, {\bf 1}, 1--19.

\bibitem[\protect\citename{Allori, }2020]{allori2020}
Allori, Valia (ed). 2020.
\newblock {\em Statistical Mechanics and Scientific Explanation: Determinism,
  indeterminism and laws of nature}.
\newblock World Scientific.

\bibitem[\protect\citename{Arntzenius, }1993]{arntzenius1993}
Arntzenius, Frank. 1993.
\newblock The Classical Failure to Account for Electromagnetic Arrows of Time.
\newblock {\em Pages  29--48 of:} Horowitz, T., \& Janis, A. (eds), {\em
  Scientific Failure}.
\newblock Rowman and Littlefield.

\bibitem[\protect\citename{Arntzenius \& Greaves, }2009]{arntzenius2009}
Arntzenius, Frank, \& Greaves, Hilary. 2009.
\newblock Time Reversal in Classical Electromagnetism.
\newblock {\em The British Journal for the Philosophy of Science}, {\bf 60},
  557--584.

\bibitem[\protect\citename{Atkinson, }2006]{atkinson2006}
Atkinson, David. 2006.
\newblock Does Quantum Electrodynamics have an Arrow of Time?
\newblock {\em Studies in History and Philosophy of Modern Physics}, {\bf 37},
  528--541.

\bibitem[\protect\citename{Bauer {\em et~al.}, }2013]{Bauer:2013aa}
Bauer, Gernot, Deckert, Dirk-Andr\'e, \& D\"urr, Detlef. 2013.
\newblock On the Existence of Dynamics of {W}heeler--{F}eynman
  Electromagnetism.
\newblock {\em Zeitschrift f{\"u}r angewandte Mathematik und Physik}, {\bf
  64}(4), 1087--1124.

\bibitem[\protect\citename{Bauer {\em et~al.}, }2014]{Bauer:2014wq}
Bauer, Gernot, Deckert, Dirk-Andr\'e, D{\"u}rr, Detlef, \& Hinrichs,
  G{\"u}nter. 2014.
\newblock On Irreversibility and Radiation in Classical Electrodynamics of
  Point Particles.
\newblock {\em Journal of Statistical Physics}, {\bf 154}(1), 610--622.

\bibitem[\protect\citename{Brown {\em et~al.}, }2009]{Brown:2009aa}
Brown, Harvey~R., Myrvold, Wayne, \& Uffink, Jos. 2009.
\newblock Boltzmann's {H}-theorem, its Discontents, and the Birth of
  Statistical Mechanics.
\newblock {\em Studies in History and Philosophy of Modern Physics}, {\bf
  40}(2), 174--91.

\bibitem[\protect\citename{Callender, }1999]{callender1999}
Callender, Craig. 1999.
\newblock Reducing Thermodynamics to Statistical Mechanics: The Case of
  Entropy.
\newblock {\em The Journal of Philosophy}, {\bf 96}(7), 348--373.

\bibitem[\protect\citename{Carroll, }2010]{carroll2010}
Carroll, Sean~M. 2010.
\newblock {\em From Eternity to Here}.
\newblock Dutton.

\bibitem[\protect\citename{Chen, }2020]{chen2020}
Chen, Eddy~Keming. 2020.
\newblock Time’s Arrow in a Quantum Universe: On the Status of Statistical
  Mechanical Probabilities.
\newblock {\em In:} Allori, V. (ed), {\em Statistical Mechanics and Scientific
  Explanation: Determinism, indeterminism and laws of nature}.
\newblock World Scientific.

\bibitem[\protect\citename{Chen, }2022a]{Chen:2020tv}
Chen, Eddy~Keming. 2022a.
\newblock From Time Asymmetry to Quantum Entanglement: The {H}umean
  unification.
\newblock {\em No{\^u}s}, {\bf 56}(1).

\bibitem[\protect\citename{Chen, }2022b]{Chen:2021uw}
Chen, Eddy~Keming. 2022b.
\newblock Fundamental Nomic Vagueness.
\newblock {\em Philosophical Review}, {\bf 131}(1), 1--49.

\bibitem[\protect\citename{Chen, }2023]{chenF}
Chen, Eddy~Keming. 2023.
\newblock The Past Hypothesis and the Nature of Physical Laws.
\newblock {\em In:} Loewer, B., Winsberg, E., \& Weslake, B. (eds), {\em The
  Probability Map of the Universe: Essays on David Albert's Time and Chance}.
\newblock Harvard University Press.

\bibitem[\protect\citename{Davies, }1977]{davies}
Davies, Paul C.~W. 1977.
\newblock {\em The Physics of Time Asymmetry}.
\newblock University of California Press.

\bibitem[\protect\citename{Dirac, }1938]{Dirac:1938aa}
Dirac, Paul A.~M. 1938.
\newblock Classical Theory of Radiating Electrons.
\newblock {\em Proceedings of the Royal Society A, Mathematical and Physical
  Sciences}, {\bf 167}(929), 148--169.

\bibitem[\protect\citename{Earman, }2011]{earman2011}
Earman, John. 2011.
\newblock Sharpening the Electromagnetic Arrow(s) of Time.
\newblock {\em Pages  485--527 of:} Callender, C. (ed), {\em The Oxford
  Handbook of Philosophy of Time}.
\newblock Oxford University Press.

\bibitem[\protect\citename{Einstein, }1909]{einstein1909}
Einstein, Albert. 1909.
\newblock \"{U}ber die Entwicklung unserer Anschauung \"{u}ber das Wesen und
  die Konstitution der Strahlung.
\newblock {\em Physikalische Zeitschrift}, {\bf 10}(22), 817--825.

\bibitem[\protect\citename{Feynman, }1965]{feynman1965}
Feynman, Richard~P. 1965.
\newblock {\em Nobel Lecture: The Development of the Space-Time View of Quantum
  Electrodynamics}.
\newblock Available at
  $\langle$\url{https://www.nobelprize.org/prizes/physics/1965/feynman/lecture/}$\rangle$.

\bibitem[\protect\citename{Feynman {\em et~al.}, }1964]{feynman2}
Feynman, Richard~P., Leighton, Robert~B., \& Sands, Matthew. 1964.
\newblock {\em The Feynman Lectures on Physics}.
\newblock  Vol. II.
\newblock Addison-Wesley Publishing Company.

\bibitem[\protect\citename{Forgione, }2020]{Forgione:2020aa}
Forgione, Marco. 2020.
\newblock The Philosophical Underpinning of the Absorber Theory of Radiation.
\newblock {\em Studies in History and Philosophy of Modern Physics}, {\bf 72},
  91--106.

\bibitem[\protect\citename{Frigg, }2008]{frigg2008}
Frigg, Roman. 2008.
\newblock A Field Guide to Recent Work on the Foundations of Statistical
  Mechanics.
\newblock {\em Pages  99--196 of:} Rickles, D. (ed), {\em The Ashgate Companion
  to Contemporary Philosophy of Physics}.
\newblock Ashgate.

\bibitem[\protect\citename{Frigg \& Werndl, }2019]{friggwerndl2019}
Frigg, Roman, \& Werndl, Charlotte. 2019.
\newblock Statistical Mechanics: A Tale of Two Theories.
\newblock {\em The Monist}, {\bf 102}, 424--438.

\bibitem[\protect\citename{Frisch, }2000]{frisch2000}
Frisch, Mathias. 2000.
\newblock (Dis-)Solving the Puzzle of the Arrow of Radiation.
\newblock {\em The British Journal for the Philosophy of Science}, {\bf 51}(3),
  381--410.

\bibitem[\protect\citename{Frisch, }2005]{frisch2005}
Frisch, Mathias. 2005.
\newblock {\em Inconsistency, Asymmetry, and Non-Locality}.
\newblock Oxford University Press.

\bibitem[\protect\citename{Frisch, }2015]{frisch2015}
Frisch, Mathias. 2015.
\newblock Why Things Happen.
\newblock {\em Aeon}.
\newblock
  \url{https://aeon.co/essays/could-we-explain-the-world-without-cause-and-effect}.

\bibitem[\protect\citename{Frisch \& Pietsch, }2016]{frischpietsch2016}
Frisch, Mathias, \& Pietsch, Wolfgang. 2016.
\newblock Reassessing the Ritz-Einstein Debate on the Radiation Asymmetry in
  Classical Electrodynamics.
\newblock {\em Studies in History and Philosophy of Modern Physics}, {\bf 55},
  13--23.

\bibitem[\protect\citename{Goldstein, }2001]{goldstein2001}
Goldstein, Sheldon. 2001.
\newblock Boltzmann's Approach to Statistical Mechanics.
\newblock {\em Pages  39--54 of:} Bricmont, J., Ghirardi, G., D\"{u}rr, D.,
  Petruccione, F., Galavotti, M.~C., \& Zanghi, N. (eds), {\em Chance in
  Physics: Foundation and Perspectives}.
\newblock Springer.

\bibitem[\protect\citename{Goldstein, }2012]{goldstein2012}
Goldstein, Sheldon. 2012.
\newblock Typicality and Notions of Probability in Physics.
\newblock {\em Pages  59--71 of:} Ben-Menahem, Y., \& Hemmo, M. (eds), {\em
  Probability in Physics}.
\newblock Springer.

\bibitem[\protect\citename{Goldstein {\em et~al.}, }2020]{goldstein2020}
Goldstein, Sheldon, Lebowitz, Joel~L., Tumulka, Roderich, \& Zangh\`i, Nino.
  2020.
\newblock Gibbs and Boltzmann Entropy in Classical and Quantum Mechanics.
\newblock {\em Pages  519--581 of:} Allori, Valia (ed), {\em Statistical
  Mechanics and Scientific Explanation}.
\newblock World Scientific.

\bibitem[\protect\citename{Griffiths, }2013]{griffiths}
Griffiths, David~J. 2013.
\newblock {\em Introduction to Electrodynamics}. 4th edn.
\newblock Prentice Hall.

\bibitem[\protect\citename{Hartenstein \& Hubert, }2021]{hartensteinhubert}
Hartenstein, Vera, \& Hubert, Mario. 2021.
\newblock When Fields Are Not Degrees of Freedom.
\newblock {\em The British Journal for the Philosophy of Science}, {\bf 72}(1),
  245--275.

\bibitem[\protect\citename{Hartle, }2005]{hartle2005}
Hartle, James~B. 2005.
\newblock The Physics of Now.
\newblock {\em American Journal of Physics}, {\bf 73}(2), 101--109.

\bibitem[\protect\citename{Hogarth, }1962]{Hogarth:1962uq}
Hogarth, J.~E. 1962.
\newblock Cosmological Considerations of the Absorber Theory of Radiation.
\newblock {\em Proceedings of the Royal Society of London. Series A,
  Mathematical and Physical Sciences}, {\bf 267}(1330), 365--383.

\bibitem[\protect\citename{Hoyle \& Narlikar, }1995]{Hoyle:1995aa}
Hoyle, F., \& Narlikar, J.~V. 1995.
\newblock Cosmology and Action-at-a-Distance Electrodynamics.
\newblock {\em Reviews of Modern Physics}, {\bf 67}(Jan), 113--155.

\bibitem[\protect\citename{Hubert, }2021]{Hubert:2020aa}
Hubert, Mario. 2021.
\newblock Reviving Frequentism.
\newblock {\em Synthese}, {\bf 199}, 5255--5284.

\bibitem[\protect\citename{Jackson, }1999]{jackson}
Jackson, John~D. 1999.
\newblock {\em Classical Electrodynamics}. 3rd edn.
\newblock Wiley.

\bibitem[\protect\citename{Jefimenko, }2000]{jefimenko2000}
Jefimenko, Oleg~D. 2000.
\newblock {\em Causality, Electromagnetic Induction, and Gravitation}. 2nd edn.
\newblock Electret Scientific Company.

\bibitem[\protect\citename{Kiessling, }2011]{Kiessling:2011ab}
Kiessling, Michael K.-H. 2011.
\newblock {\em On the Motion of Point Charges Coupled to the
  {Maxwell--Born--Infeld} Fields}.
\newblock Lecture notes for the 457.\ WE-Heraeus-Seminar in Bad Honnef.

\bibitem[\protect\citename{Kuhn, }1978]{Kuhn:1978aa}
Kuhn, Thoms~S. 1978.
\newblock {\em Black-Body Theory and the Quantum Discontinuity 1894--1912}.
\newblock Chicago: The University of Chicago Press.

\bibitem[\protect\citename{Lange, }2002]{lange}
Lange, Marc. 2002.
\newblock {\em An Introduction to the Philosophy of Physics: Locality, Energy,
  Fields, and Mass}.
\newblock Blackwell.

\bibitem[\protect\citename{Lazarovici, }2018]{lazarovici2018}
Lazarovici, Dustin. 2018.
\newblock Against Fields.
\newblock {\em European Journal for Philosophy of Science}, {\bf 8}, 145--170.

\bibitem[\protect\citename{Loewer, }2012a]{Loewer:2012aa}
Loewer, Barry. 2012a.
\newblock The Emergence of Time's Arrows and Special Science Laws from Physics.
\newblock {\em Interface Focus}, {\bf 2}, 13--19.

\bibitem[\protect\citename{Loewer, }2012b]{Loewer:2012ab}
Loewer, Barry. 2012b.
\newblock Two Accounts of Laws and Time.
\newblock {\em Philosophical Studies}, {\bf 160}(1), 115--37.

\bibitem[\protect\citename{Loewer, }2020]{Loewer:2020aa}
Loewer, Barry. 2020.
\newblock The Mentaculus Vision.
\newblock {\em Pages  3--29 of:} Allori, Valia (ed), {\em Statistical Mechanics
  and Scientific Explanation}.
\newblock World Scientific.

\bibitem[\protect\citename{Malament, }2004]{malament2004}
Malament, David~B. 2004.
\newblock On the Time Reversal Invariance of Classical Electromagnetic Theory.
\newblock {\em Studies In History and Philosophy of Science Part B: Studies In
  History and Philosophy of Modern Physics}, {\bf 35}(2), 295--315.

\bibitem[\protect\citename{Maudlin, }2007]{Maudlin:2007ad}
Maudlin, Tim. 2007.
\newblock On the Passing of Time.
\newblock {\em Chap. 4, pages  104--142 of:} {\em The Metaphysics Within
  Physics}.
\newblock Oxford University Press.

\bibitem[\protect\citename{Maudlin, }2018]{Maudlin:2018aa}
Maudlin, Tim. 2018.
\newblock Ontological Clarity via Canonical Presentation: Electromagnetism and
  the {A}haronov--{B}ohm Effect.
\newblock {\em Entropy}, {\bf 20}(6), 1--21.

\bibitem[\protect\citename{Maudlin, }2020]{Maudlin:2019ab}
Maudlin, Tim. 2020.
\newblock The Grammar of Typicality.
\newblock {\em Chap. 7, pages  231--251 of:} Allori, Valia (ed), {\em
  Statistical Mechanics and Scientific Explanation: Determinism, Indeterminism
  and Laws of Nature}.
\newblock World Scientific.

\bibitem[\protect\citename{Mundy, }1989]{mundy1989}
Mundy, Brent. 1989.
\newblock Distant Action in Classical Electromagnetic Theory.
\newblock {\em The British Journal for the Philosophy of Science}, {\bf 40}(1),
  39--68.

\bibitem[\protect\citename{Myrvold, }2021]{Myrvold:2021tx}
Myrvold, Wayne~C. 2021.
\newblock {\em Beyond Chance and Credence}.
\newblock Oxford University Press.

\bibitem[\protect\citename{North, }2003]{north2003}
North, Jill. 2003.
\newblock Understanding the Time-Asymmetry of Radiation.
\newblock {\em Philosophy of Science: Proceedings of the 2002 Biennial Meeting
  of the Philosophy of Science Association}, {\bf 70}(5), 1086--1097.

\bibitem[\protect\citename{North, }2011]{north2011}
North, Jill. 2011.
\newblock Time in Thermodynamics.
\newblock {\em Pages  312--350 of:} Callender, C. (ed), {\em The Oxford
  Handbook of Philosophy of Time}.
\newblock Oxford University Press.

\bibitem[\protect\citename{Penrose \& Percival, }1962]{penrosepercival}
Penrose, O., \& Percival, I.~C. 1962.
\newblock The Direction of Time.
\newblock {\em Proceedings of the Physical Society}, {\bf 79}, 605--616.

\bibitem[\protect\citename{Penrose, }2001]{Penrose:2001aa}
Penrose, Oliver. 2001.
\newblock The Direction of Time.
\newblock {\em In:} Bricmont, Jean, D{\"u}rr, Detlef, Galavotti, Maria~Carla,
  Ghirardi, Giancarlo, \& Petruccione, Francesco (eds), {\em Chance in Physics:
  Foundations and Perspectives}.
\newblock Heidelberg: Springer.

\bibitem[\protect\citename{Penrose, }1979]{penrose1979}
Penrose, Roger. 1979.
\newblock Singularities and Time-Asymmetry.
\newblock {\em Chap. 12, pages  581--638 of:} Hawking, S.~W., \& Israel, W.
  (eds), {\em General Relativity: An Einstein Centenary Survey, Part I}.
\newblock Cambridge University Press.

\bibitem[\protect\citename{Pietsch, }2012]{pietsch2012}
Pietsch, Wolfgang. 2012.
\newblock Hidden Underdetermination: A Case Study in Classical Electrodynamics.
\newblock {\em International Studies in the Philosophy of Science}, {\bf
  26}(2), 125--151.

\bibitem[\protect\citename{Popper, }1956]{popper1956}
Popper, Karl. 1956.
\newblock The Arrow of Time.
\newblock {\em Nature}, {\bf 177}, 538.

\bibitem[\protect\citename{Popper, }1958]{popper1958}
Popper, Karl. 1958.
\newblock Irreversible Processes in Physical Theory.
\newblock {\em Nature}, {\bf 181}, 402--403.

\bibitem[\protect\citename{Price, }1996]{price1996}
Price, Huw. 1996.
\newblock {\em Time's Arrow and Archimedes' Point}.
\newblock Oxford University Press.

\bibitem[\protect\citename{Price, }2004]{price2004}
Price, Huw. 2004.
\newblock On the Origins of the Arrow of Time: Why there is Still a Puzzle
  about the Low-Entropy Past.
\newblock {\em Pages  219--239 of:} Hitchcock, C. (ed), {\em Contemporary
  Debates in Philosophy of Science}.
\newblock Blackwell.

\bibitem[\protect\citename{Price, }2006]{price2006}
Price, Huw. 2006.
\newblock Recent Work on the Arrow of Radiation.
\newblock {\em Studies in History and Philosophy of Modern Physics}, {\bf 37},
  498--527.

\bibitem[\protect\citename{Rinard, }2017]{Rinard:2021vb}
Rinard, Susanna. 2017.
\newblock Imprecise Probability and Higher Order Vagueness.
\newblock {\em Res Philosophica}, {\bf 94}(2), 257--273.
\newblock 2017.

\bibitem[\protect\citename{Ritz \& Einstein, }1909]{ritzeinstein1909}
Ritz, Walther, \& Einstein, Albert. 1909.
\newblock Zum gegenw\"{a}rtigen Stand des Strahlungsproblems.
\newblock {\em Physikalische Zeitschrift}, {\bf 10}(9), 323--324.
\newblock English translation: \citet{ritzeinstein1990}.

\bibitem[\protect\citename{Ritz \& Einstein, }1990]{ritzeinstein1990}
Ritz, Walther, \& Einstein, Albert. 1990.
\newblock On the Present Status of the Radiation Problem.
\newblock {\em Page  376 of:} Beck, A. (ed), {\em The Collected Papers of
  Albert Einstein, Volume 2: The Swiss Years: Writings, 1900-1909 (English
  translation supplement)}.
\newblock Pinceton University Press.
\newblock Translated by Anna Beck from \citet{ritzeinstein1909}.

\bibitem[\protect\citename{Roberts, }2021]{roberts2021}
Roberts, Bryan. 2021.
\newblock Time Reversal.
\newblock {\em Pages  605--619 of:} Knox, E., \& Wilson, A. (eds), {\em The
  Routledge Companion to Philosophy of Physics}.
\newblock Routledge.

\bibitem[\protect\citename{Rohrlich, }2006]{Rohrlich:2006aa}
Rohrlich, F. 2006.
\newblock Time in Classical Electrodynamics.
\newblock {\em American Journal of Physics}, {\bf 74}(4), 313--315.

\bibitem[\protect\citename{Rohrlich, }1999]{rohrlich1999}
Rohrlich, Fritz. 1999.
\newblock Classical Self-Force.
\newblock {\em Physical Review D}, {\bf 60}, 084017.

\bibitem[\protect\citename{Rohrlich, }2000]{rohrlich2000}
Rohrlich, Fritz. 2000.
\newblock Causality and the Arrow of Classical Time.
\newblock {\em Studies in History and Philosophy of Modern Physics}, {\bf
  31}(1), 1--13.

\bibitem[\protect\citename{Rohrlich, }2002]{rohrlich2002}
Rohrlich, Fritz. 2002.
\newblock Causality, the Coulomb Field, and Newton's Law of Gravitation.
\newblock {\em American Journal of Physics}, {\bf 70}, 411--414.

\bibitem[\protect\citename{Rohrlich, }2007]{rohrlich}
Rohrlich, Fritz. 2007.
\newblock {\em Classical Charged Particles}. 3rd edn.
\newblock World Scientific.

\bibitem[\protect\citename{Schot, }1992]{Schot:1992aa}
Schot, Steven~H. 1992.
\newblock Eighty Years of {S}ommerfeld's Radiation Condition.
\newblock {\em Historia Mathematica}, {\bf 19}(4), 385--401.

\bibitem[\protect\citename{Schwinger {\em et~al.}, }1998]{Schwinger:1998uq}
Schwinger, Julian, Deraad, Lester~L., Milton, Kimball~A., Tsai, Wu-Yang, \&
  Norton, Joyce. 1998.
\newblock {\em Classical Electrodynamics}.
\newblock Reading, MA: Perseus Books.

\bibitem[\protect\citename{Sebens, }2020]{positrons}
Sebens, Charles~T. 2020.
\newblock Putting Positrons into Classical Dirac Field Theory.
\newblock {\em Studies in History and Philosophy of Modern Physics}, {\bf 70},
  8--18.

\bibitem[\protect\citename{Sebens, }2022a]{potentialenergy}
Sebens, Charles~T. 2022a.
\newblock The Disappearance and Reappearance of Potential Energy in Classical
  and Quantum Electrodynamics.
\newblock {\em Foundations of Physics}, {\bf 52}(113), 1--30.

\bibitem[\protect\citename{Sebens, }2022b]{fundamentalityoffields}
Sebens, Charles~T. 2022b.
\newblock The Fundamentality of Fields.
\newblock {\em Synthese}, {\bf 200}(5), 380.

\bibitem[\protect\citename{Sebens, }2022c]{gravitationalfield}
Sebens, Charles~T. 2022c.
\newblock The Mass of the Gravitational Field.
\newblock {\em The British Journal for the Philosophy of Science}, {\bf 73}(1),
  211--248.

\bibitem[\protect\citename{Sommerfeld, }1949]{Sommerfeld:1949aa}
Sommerfeld, Arnold. 1949.
\newblock {\em Partial Differential Equations in Physics}.
\newblock New York: Academic Press.

\bibitem[\protect\citename{Struyve, }forthcoming]{struyve2020}
Struyve, Ward. forthcoming.
\newblock Time-Reversal Invariance and Ontology.
\newblock {\em The British Journal for the Philosophy of Science}.

\bibitem[\protect\citename{Uffink, }2007]{uffink2007}
Uffink, Jos. 2007.
\newblock Compendium of the Foundations of Classical Statistical Physics.
\newblock {\em Pages  923--1074 of:} Butterfield, J., \& Earman, J. (eds), {\em
  Handbook of the Philosophy of Science: Philosophy of Physics, Part B}.
\newblock Elsevier.

\bibitem[\protect\citename{Wald, }2022]{wald2022}
Wald, Robert~M. 2022.
\newblock {\em Advanced Classical Electromagnetism}.
\newblock Princeton University Press.

\bibitem[\protect\citename{Wallace, }2015]{wallace2015}
Wallace, David. 2015.
\newblock The Quantitative Content of Statistical Mechanics.
\newblock {\em Studies in History and Philosophy of Modern Physics}, {\bf 52},
  285--293.

\bibitem[\protect\citename{Wallace, }2023]{wallace2011}
Wallace, David. 2023.
\newblock The Logic of the Past Hypothesis.
\newblock {\em In:} Loewer, B., Winsberg, E., \& Weslake, B. (eds), {\em The
  Probability Map of the Universe: Essays on David Albert's Time and Chance}.
\newblock Harvard University Press.

\bibitem[\protect\citename{Wheeler \& Feynman, }1945]{Wheeler:1945aa}
Wheeler, John~Archibald, \& Feynman, Richard~Phillips. 1945.
\newblock Interaction with the Absorber as the Mechanism of Radiation.
\newblock {\em Reviews of Modern Physics}, {\bf 17}(2-3), 157--181.

\bibitem[\protect\citename{Wheeler \& Feynman, }1949]{Wheeler:1949aa}
Wheeler, John~Archibald, \& Feynman, Richard~Phillips. 1949.
\newblock Classical Electrodynamics in Terms of Direct Interparticle Action.
\newblock {\em Reviews of Modern Physics}, {\bf 21}(3), 425--433.

\bibitem[\protect\citename{Zeh, }2007]{zeh2007}
Zeh, H.~Dieter. 2007.
\newblock {\em The Physical Basis of the Direction of Time}. 5th edn.
\newblock Springer.

\end{thebibliography}
\end{document}